\documentclass[fleqn,usenatbib]{mnras}

\usepackage{caption}
\usepackage{subcaption}

\usepackage[T1,T4]{fontenc}
\usepackage{ae,aecompl}

\DeclareRobustCommand{\VAN}[3]{#2}
\let\VANthebibliography\thebibliography
\def\thebibliography{\DeclareRobustCommand{\VAN}[3]{##3}\VANthebibliography}

\usepackage{graphicx}	% Including figure files
\usepackage{amsmath}	% Advanced maths commands
\usepackage{txfonts}
\usepackage[usenames,dvipsnames]{xcolor}
\usepackage{xfrac}
\usepackage{url}
\usepackage{float}
\usepackage{pdflscape}
\usepackage{soul}
\usepackage{enumitem}
\usepackage[normalem]{ulem}
\usepackage[symbol]{footmisc}

\newcounter{daggerfootnote}

\title[\texttt{Simba}-\texttt{C}: properties of the IGrM]{\texttt{Simba}-\texttt{C}: the evolution of the thermal and chemical properties in the intragroup medium}

\author[R. T. Hough et al.]{
Renier T. Hough,$^{1}$\thanks{E-mail: renierht@gmail.com}
Zhiwei Shao,$^{2}$
Weiguang Cui,$^{3,4,5}$
S. Ilani Loubser,$^{1, 6}$
\stepcounter{daggerfootnote}Arif Babul,$^{7}$\thanks{Infosys  Visiting Chair Professor, Indian Institute of Science, Bangalore 560012, India}
Romeel Dav\'e,$^{3,8}$
 \newauthor Douglas Rennehan$^{7, 9}$and Chiaki Kobayashi$^{10}$
\\
$^{1}$Centre for Space Research, North-West University, Potchefstroom 2520, South Africa\\
$^{2}$Department of Astronomy, School of Physics and Astronomy, and Shanghai Key Laboratory for Particle Physics and Cosmology,\\ Shanghai Jiao Tong University, Shanghai 200240, China
$^{3}$School of Physics and Astronomy, University of Edinburgh, Edinburgh EH9 3HJ, UK\\
$^{4}$Departamento de Física Teórica, M-8, Universidad Autónoma de Madrid, Cantoblanco 28049, Madrid, Spain\\
$^{5}$Centro de Investigación Avanzada en Física Fundamental (CIAFF), Universidad Aut\'{o}noma de Madrid, Cantoblanco, 28049 Madrid, Spain\\
$^{6}$National Institute for Theoretical and Computational Sciences (NITheCS), Potchefstroom 2520, South Africa\\
$^{7}$Department of Physics \& Astronomy, University of Victoria, Victoria BC V8P 5C2, Canada\\
$^{8}$Department of Physics and Astronomy, University of the Western Cape, Bellville, 7535, South Africa\\
$^{9}$Center for Computational Astrophysics, Flatiron Institute, 162 5th Ave, New York, NY 10010, USA\\
$^{10}$Centre for Astrophysics Research, Department of Physics, Astronomy and Mathematics, University of Hertfordshire, Hatfield AL10 9AB, UK\\
}

\date{Accepted XXX. Received YYY; in original form ZZZ}

\pubyear{2023}

\begin{document}
\label{firstpage}
\pagerange{\pageref{firstpage}--\pageref{lastpage}}

\maketitle

\begin{abstract}
The newly updated \texttt{GIZMO} and \texttt{Simba} based simulation, \texttt{Simba-C}, with its new stellar feedback, chemical enrichment, and recalibrated AGN feedback, allows for a detailed study of the intragroup medium X-ray properties. We discuss the impact of various physical mechanisms, e.g. stellar and AGN feedback, and chemical enrichment, on the composition and the global scaling relations of nearby galaxy groups. We also study the evolution ($z=2$ to $0$) of the global properties for the $1\,\mathrm{keV}$ temperature groups. \texttt{Simba-C} shows improved consistent matching with the observations of all X-ray scaling relations compared to \texttt{Simba}. It is well known that AGN feedback has a significant influence on $L_{X,0.5-2.0}-T_{spec,corr}$, $S_{500/2500}-T_{spec,corr}$, and gas mass fractions, with our \texttt{Simba-C} results consistent with it. Our recalibrated AGN feedback strength also showed an additional improvement in gas entropy, which now aligns with CLoGS observations. The updated stellar feedback and chemical enrichment model is shown to play an important role in our understanding of the chemical abundance ratios and their evolution within galaxy groups. In particular, we find that \texttt{Simba-C} produces an increase in the amount of heavier elements (specifically Si and Fe) relative to O, compared to \texttt{Simba}.
\end{abstract}

\begin{keywords}
galaxies: haloes -- galaxies: abundances -- galaxies: evolution -- galaxies: formation -- galaxies: stellar content -- X-rays: galaxies %\url{https://static.primary.prod.gcms.the-infra.com/static/site/mnras/document/MNRAS\%20Keywords_November\%202022.pdf?node=d966aa6b52fb4ac858b7}
\end{keywords}

%%%%%%%%%%%%%%%%%%%%%%%%%%%%%%%%%%%%%%%%%%%%%%%%%%

%%%%%%%%%%%%%%%%% BODY OF PAPER %%%%%%%%%%%%%%%%%%

\section{Introduction}

Galaxy groups contain most of the bound baryon content in the Universe \citep{Fukugita1998, Lovisari2021b}, including more than half of all galaxies \citep{Eke2006, Lovisari2021a}. A significant fraction of these baryons exist in the form of hot diffuse gas \citep{Finoguenov2003, McNamara2007}, which can be studied using X-ray observations \citep[e.g.][and references therein]{Ponman2003, McNamara2007, Liang2016, OSullivan2017, Gastaldello2021, Lovisari2021a}. Such observations reveal the impact of various processes that occur within galaxies, such as star formation, stellar nucleosynthesis, stellar and active galactic nuclei (AGN) feedback, and galactic outflows \citep[e.g.][]{Babul2002, Pakmor2023}. Therefore, groups are excellent laboratories for studying the impact that galaxies have on their surroundings via baryon cycling processes \citep{Renzini1997, Finoguenov2003, McCarthy2008, Dave2008, Liang2016, Oppenheimer2021, Saeedzadeh2023, Loubser2024}.

From X-ray observations, the following three group properties have attracted the most attention:

(i) Entropy of the hot diffuse gas ($S=k_{\mathrm{B}}T_{\mathrm{e}}/n_{\mathrm{e}}^{2/3}$) within $R_{500}$ -- This quantity contains the time-integrated history of the heating and cooling that the gas has experienced, including non-gravitational heating induced by stellar-powered galactic outflows and/or AGNs \citep{Lewis2000, Sijacki2008,  LeBrun2014}. Entropy offers a more precise time-integrated representation of the energy flow within these groups compared to temperature or density measurements alone \citep{Balogh1999, Babul2002}. % Babul2002, Borgani2004, Dave2008, Everett2008, McCarthy2008, McCarthy2010, McCarthy2011, Puchwein2008, Teyssier2011, Hopkins2012, Short2013, Socrates2008, Planelles2014, Zhang2014

(ii) Hot gas fractions in central regions -- The hot gas fraction in the central region of groups, i.e. within $R_{2500}$, is considerably lower than that found in massive clusters \citep{Balogh1999, Vikhlinin2006, Gastaldello2007, Sun2009}. This discrepancy could arise from various processes, such as the non-gravitational heating of the gas described in (i) driving its expulsion \citep[e.g.][]{Balogh1999, Dave2008, Liang2016}, or the depletion of gas due to efficient cooling~\citep[e.g.][]{Voit2001}. Solely the latter would lead to significantly higher stellar mass fractions than those observed \citep{Dave2002}, but when combined with feedback processes, it could become an important contributing factor.

(iii) Metal content of the hot diffuse gas -- Heavier elements originate from stellar nucleosynthesis, primarily through core-collapse supernova explosions of massive stars (SNe II) and thermonuclear detonations of accreting white dwarf stars (SNe Ia), and are dispersed into the broader environment via galactic outflows \citep{Dave2008} and/or ram pressure stripping \citep{Domainko2006}. It is believed that SNe II seeded the Universe with metals during the early stages of galaxy formation, whereas SNe Ia are associated with the late stages of stellar evolution of smaller stars, dominating metal production over longer periods \citep{McNamara2007}. Consequently, X-ray observations of the relative abundances of the by-products of SNe II vs. SNIa enable the constraint of galaxy groups' star formation histories \citep{McNamara2007}. Furthermore, larger groups and clusters are observed to have an iron abundance of approximately $[\mathrm{Fe/H}] \sim 0.3$ \citep{Edge1991, Peterson2003, DeGrandi2004, DePlaa2007}, indicating that a significant fraction of the metals produced escape the interstellar medium (ISM), potentially with a considerable fraction of the metals having been ejected prior to the formation of the groups themselves \citep{Oppenheimer2012}. Recent studies (e.g. \citealt{Appleby2021, Saeedzadeh2023}) consider strong outflows/winds as the most plausible candidate for enrichment, although ram pressure stripping also contributes \citep{Saeedzadeh2023}. Once the metals leave the ISM, they are further redistributed by outflow-driven turbulence \citep{Rennehan2019,Bennett2020, Lochhaas2020, Rennehan2021, Li2023}.

From these three properties, it becomes evident that feedback-driven outflows play a pivotal role in the formation and evolution of galaxies within group environments. These winds are more ubiquitous at high redshifts but are now also locally observable \citep{Martin2005, Sturm2011, Veilleux2013, Turner2014}. It is suggested that outflows arise from either stellar processes (e.g. SNe explosions) or AGNs and possess the capacity to alter the physical properties of their environments, since the feedback energy is comparable to the binding energies of groups \citep{Lovisari2021a}. Specifically, AGN-powered winds originate as high-velocity outflows on parsec scales at the centre of galaxies and have a profound impact on the gas in the central ($\sim 1\, \mathrm{ kpc}$) region of their host galaxies \citep{Sturm2011, Veilleux2013, Villar2014}.  Consequently, it is interesting to ask questions about their impact on the large-scale group environment \citep{Yang2024}. %Martin2006, OSullivan2012, Bradshaw2013, Sell2014, Villar2014, Williams2015

Stellar-powered outflows, driven by SNe energy and momentum injection, along with photoheating and radiation pressure from massive stars \citep{Murray2005, Murray2010, Krumholz2013}, exhibit remarkable efficacy in the removal of metal-enriched ISM gas \citep{Taylor2015}. This effectiveness stems from the launch sites of the winds being embedded in star-forming regions throughout the ISM, with metal-enriched winds observed to reach velocities exceeding the escape velocity with mass outflow rates comparable to the galaxy's star formation rate (SFR).  Simulations such as the Feedback In Realistic Environments (FIRE) simulations \citep{Hopkins2014} and the Numerical Investigation of a Hundred Astrophysical Objects (NIHAO) simulations \citep{Wang2015} have demonstrated that stellar feedback is capable of launching powerful galaxy-wide winds with substantial mass loading, transporting metals into the intragroup medium (IGrM) and beyond. The diverse underlying galaxy processes combined with the galactic outflows generate dynamic changes observable in group halo properties, particularly in the hot X-ray gas. Such observations provide an opportunity to constrain the as-yet poorly understood stellar and AGN feedback mechanisms.

In this paper, we study intragroup gas properties using our latest \texttt{Simba-C} galaxy evolution simulation \citep{Hough2023}, which is based on the \texttt{Simba} galaxy formation model \citep{Dave2019}. This updated version incorporates enhancements such as an improved chemical enrichment and stellar feedback model, a reintegrated dust model, and a modified AGN feedback model. Our investigation aims to understand the influence of metal-enriched outflows on various characteristics of hot diffuse gas X-ray properties, including entropy, temperature, luminosity, mass, and the hot gas mass fractions within central regions. Using the newly updated chemical enrichment model \citep{Hough2023}, we also examine the metal content present in this hot diffuse gas, contrasting it with the approximation of instantaneous recycling of the metals model \citep{Dave2019}.

Sec. \ref{sec: Simulation methodology}, discusses the input physics of the \texttt{Simba-C} simulation, highlighting its significantly modified chemical enrichment and feedback modules. Additionally, this section also provides a brief description of how we compile our catalogue of simulated galaxy groups and outlines the distinctions between the various versions of the \texttt{Simba}/\texttt{Simba-C} simulations. In Sec. \ref{sec: scaling relations}, our focus lies on discussing the global X-ray properties of our simulated galaxy groups at $z=0$, emphasizing the three commonly discussed group X-ray scaling relations: (i) Luminosity-temperature, (ii) mass-temperature, and (iii) entropy-temperature. In Sec. \ref{sec: Baryon fractions} and \ref{sec: Metal enrichment of IGrM}, we discuss the gas mass fractions and the metal content of the IGrM, respectively. In Sec. \ref{sec: Redshift evolution}, we discuss the evolution ($z=2$ to $0$) of the $1\,\mathrm{keV}$ temperature groups for each of the various scaling relations and physical properties discussed in Sec. \ref{sec: scaling relations}. Finally, we summarize our findings and conclusions in Sec. \ref{sec: Conclusions}.

\section{Simulation methodology}\label{sec: Simulation methodology}
\subsection{\texttt{Simba-C}}\label{sec: Simba-C simulation}

In this paper, we use \texttt{Simba-C} as our main galaxy evolution model. This simulation \citep{Hough2023} is a forked version of the original \texttt{Simba} simulation \citep{Dave2019}, a comprehensive large-volume cosmological simulation utilizing the hydrodynamics+gravity solver \texttt{GIZMO} \citep{Springel2005, Hopkins2015, Hopkins2017}. This section provides an overview of the \texttt{Simba-C} simulation, with further detailed information available in \citet{Dave2019} and \citet{Hough2023} for interested readers. \texttt{GIZMO} evolves the hydrodynamic equations using the mesh-free finite-mass (MFM) method \citep{Lanson2008a, Lanson2008b, Gaburov2011}. \texttt{GIZMO} also handles shocks using a Riemann solver without artificial viscosity, while preserving the mass within each fluid element at simulation time, facilitating the simple tracking of gas flows \citep{Hopkins2015, Dave2019, Asensio2023}.

Both the \texttt{Simba} and \texttt{Simba-C} simulations utilize the \texttt{GRACKLE-3.1} library \citep{Smith2016}, to handle the radiative cooling and photoionization heating of the gas. This also includes metal cooling and the non-equilibrium evolution of the primordial elements. Within this framework, adiabatic and radiative terms evolve simultaneously during a cooling sub-time-step, ensuring a stable thermal evolution. The model also accounts for self-shielding self-consistently based on a local attenuation approximation \citep{Smith2016, Dave2019}. Star formation is modelled using an $\mathrm{H_{2}}$-based SFR, where the fraction of $\mathrm{H_{2}}$ is based on the sub-grid model that considers the metallicity and local column density of $\mathrm{H_{2}}$, based on the approach outlined in \citet{Krumholz2011}. 

\texttt{Simba-C} differentiates itself from \texttt{Simba} by adopting the \texttt{Chem5} cosmic chemical enrichment model, developed and refined in various studies, including \citet{Kobayashi2004}, \citet{Kobayashi2007}, \citet{Taylor2014}, \citet{Kobayashi2011a}, \citet{Kobayashi2020b}, and \citet{Kobayashi2020a}. The \texttt{Chem5} model is the `version-5' of a self-consistent\footnote{The model tracks the metal return following a detailed stellar evolution model with mass- and metal-dependent yields \citep{Hough2023}.} chemodynamical enrichment model, tracking all elements from Hydrogen (H) to Germanium (Ge). It accounts for a range of physical stellar feedback processes: Stellar winds, Asymptotic Giant Branch (AGB) stars, super AGB stars, Type Ia SNe, Type II SNe -- including Hypernovae (HNe), and `failed' SNe for the most massive stars. Importantly, \texttt{Chem5} does not use the instantaneous metal recycling approximation or a simplified delayed feedback model for Type Ia SNe and AGB stars. Instead, it treats each star particle as an evolving stellar population that ejects energy, mass, and metals into its nearby environment. This fundamental difference between the original \texttt{Simba} simulation and the \texttt{Simba-C} simulation leads to a more time-resolved modelling of stellar enrichment, relying on updated yields and stellar evolution models. This approach offers a more detailed and comprehensive representation of stellar enrichment dynamics. We refer interested readers to \citet{Hough2023} for a more in-depth discussion of the implementation of the \texttt{Chem5} model into the \texttt{Simba-C} simulation, as well as the comparative tests to the \texttt{Simba} simulation. A key benefit of \texttt{Simba-C} over its predecessor is its improved ability to predict observed abundance ratios. 

\texttt{Simba-C} uses \texttt{Simba}'s model for star formation-driven galactic winds, which employs decoupled two-phase winds with a mass load factor. It also uses the on-the-fly approximated friend-of-friends (FOF) finder, as described in \citet{Dave2016}, to compute various galaxy properties to which the mass loading and wind velocity are scaled. While the scaling of mass loading remains unchanged from \texttt{Simba}, there is a reduced normalization modification in the velocity scaling, with \texttt{Simba-C} utilizing 0.85 instead of the previous value of 1.6. For a detailed understanding of the rationale behind this re-calibration, refer to \citet{Hough2023}.

\texttt{Simba-C} mostly follows \texttt{Simba}'s black hole physics with some updates. Black holes are seeded when the galaxy is initially resolved ($M_{*}\gtrsim 6\times 10^{8}\,\mathrm{ M_{\odot}}$). However, the black hole accretion is suppressed exponentially by a factor of $1-\mathrm{exp}(-M_{\mathrm{BH}}/10^{6}\,\mathrm{ M_{\odot}})$, aimed at mimicking the effect of star formation suppressing BH growth in dwarf galaxies as seen in simulations as described in \citet{Angles2017b} and \citet{Hopkins2022}. At simulation time, both the BHs' dynamical mass (inherited from the parent star particle) and their physical black hole mass (set to $M_{\mathrm{BH, seed}} = 10^{4}\,\mathrm{ M_{\odot}}\,h^{-1}$ and allowed to grow via accretion) are tracked.

\texttt{Simba-C} continues to use \texttt{Simba}'s two-mode black hole accretion model. For the cool gas ($T<10^5\,\mathrm{K}$), we compute the accretion rate based on the torque-limited accretion model (limited to three times the Eddington accretion rate) presented in \citet{Hopkins2011} and \citet{Angles2017c}
\begin{equation}
\begin{split}
\dot{M}_{\mathrm{Torque}} &\equiv \epsilon_{T}f^{5/2}_{d}\bigg(\frac{M_{\mathrm{BH}}}{10^{8}\,\textup{M}_{\odot}}\bigg)^{1/6} \bigg(\frac{M_{\mathrm{enc}}(R_{0})}{10^{9}\,\textup{M}_{\odot}}\bigg)\\
    &\bigg(\frac{R_{0}}{100\,\textup{pc}}\bigg)^{-3/2}\bigg(1+\frac{f_{0}}{f_{\mathrm{gas}}}\bigg)^{-1}\textup{M}_{\odot}\,\textup{yr}^{-1}. \\
\end{split}
\end{equation}
Hot gas, meanwhile, is accreted following the standard Bondi-Hoyle-Lyttleton accretion rate (limited to the Eddington rate) presented in \citet{Bondi1952}
\begin{equation}
\begin{split}
\dot{M}_{\mathrm{Bondi}} &= \frac{4\pi G^{2}M^{2}_{\mathrm{BH}}\rho}{c_{\mathrm{s}}^{3}}.
\end{split}
\end{equation}
As in \texttt{Simba}, the total large-scale accretion rate onto each black hole is given as the sum of the two modes, taking into account the conversion of matter into radiation
\begin{equation}
    \dot{M}_\mathrm{BH} = \left(1-\eta\right) \times \left(\dot{M}_{\mathrm{Torque}} + \dot{M}_{\mathrm{Bondi}}\right),
\end{equation}
where $\eta=0.1$ is the radiative efficiency \citep{Yu2002}.

\texttt{Simba-C} uses the accretion energy to drive the black hole feedback that quenches galaxies. This is achieved through a kinetic subgrid model for black hole feedback combined with X-ray energy feedback, as described in \citet{Dave2019}. Furthermore, \texttt{Simba-C}'s AGN feedback model mimics the energy injection into the large-scale surrounding gas by using purely bipolar feedback in the angular momentum direction of the black hole accretion radius. Regarding the mass scale at which the jets are permitted, \texttt{Simba-C} sets a range between $7\times 10^{7}\,\mathrm{M_{\odot}} - 1\times 10^{8}\,\mathrm{M_{\odot}}$. Additionally, each black hole particle is effectively assigned its own jet onset mass which it retains throughout the simulation run. Another distinction from \texttt{Simba} in \texttt{Simba-C} is that the black hole maximum jet velocity boost is allowed to reach a value that scales with the halo escape velocity, rather than a constant value of $7000\,\mathrm{km/s}$ (see \citet{Hough2023} for more details).

\subsection{Dust integration and results}

The sole difference between the \texttt{Simba-C} simulation results presented in \citet{Hough2023} and the simulation utilized in this article is the inclusion of \texttt{Simba}'s dust model. The model was intentionally omitted in \citet{Hough2023}, due to its possible influence on the metal content in a simulation dedicated to testing the new metal enrichment model. 

For this paper, we re-introduce \texttt{Simba}'s dust model to strengthen the precision of our representations of a physical system. As described in \citet{Dave2019}, the dust is passively advected following the gas particles. Dust is produced by the condensation of metals from the ejecta of SNe and AGB stars. \texttt{Simba} uses fixed dust condensation efficiencies of $\delta^{\mathrm{AGB}}_{i,\mathrm{dust}} = 0.2$ and $\delta^{\mathrm{SNII}}_{i,\mathrm{dust}} = 0.15$, based on theoretical models from \citet{Ferrarotti2006} and \citet{Remy2014} to match the low metallicity end of the observed dust-to-gas mass ratios. It should be noted that Type Ia SNe condensation is omitted due to its low impact on dust production \citep{Nozawa2011, Dwek2016, Gioannini2017}. Dust can grow, be destroyed, and undergo thermal sputtering \citep{Li2020}. However, it is essential to note that the dust model is still applied only to the original 11 elements that were tracked in \texttt{Simba}. The additional elements introduced in \texttt{Simba-C} are assumed to not participate in the formation of dust, as they constitute a very small fraction of the total metal mass.

\begin{figure}
\includegraphics[width=\columnwidth]{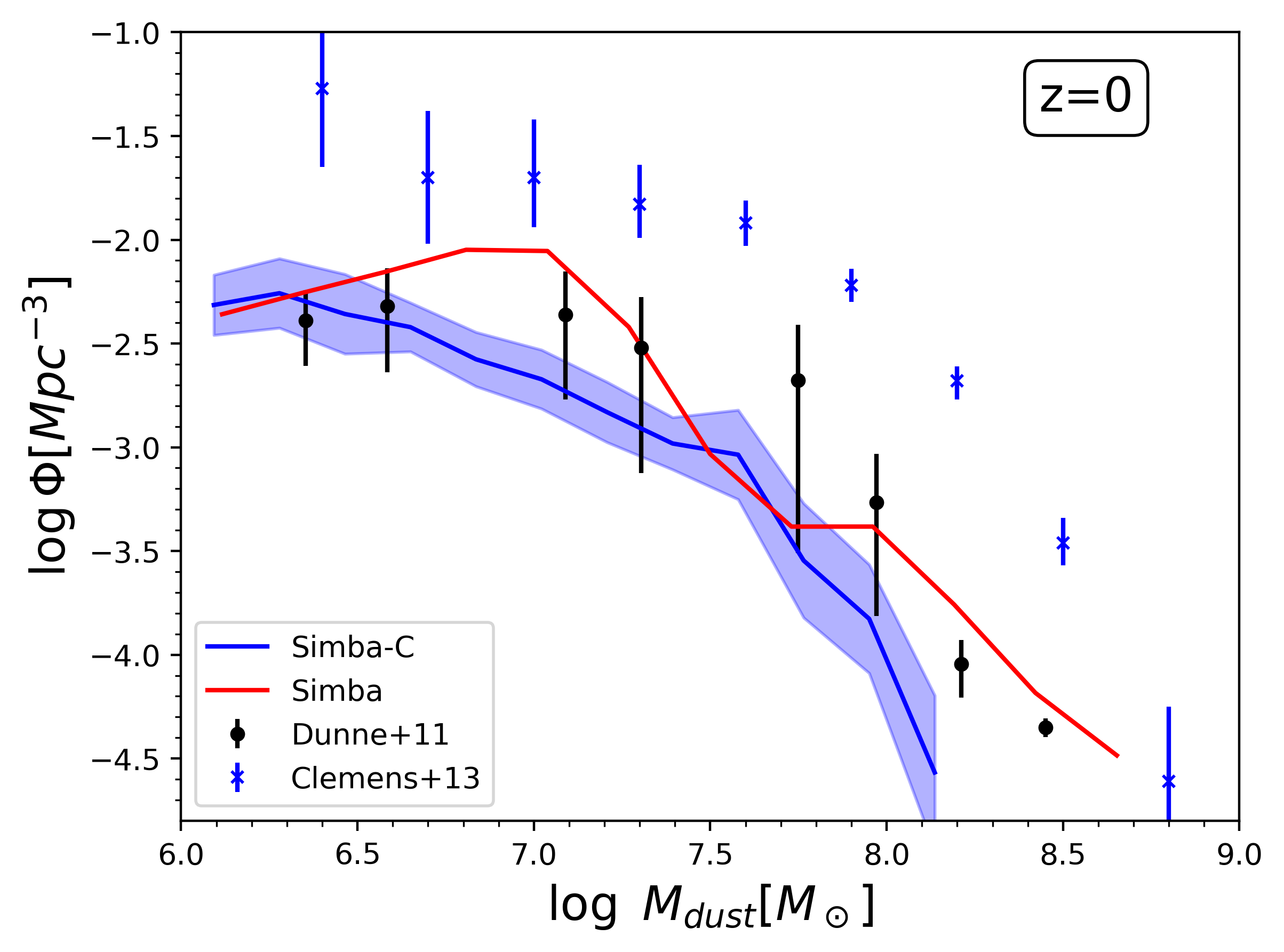}
    \caption{Comparison of the dust mass function between the \texttt{Simba}-\texttt{C} simulation and the published version of \texttt{Simba} at $z=0$, compared to observations from \citet{Dunne2011} and \citet{Clemens2013}. The \texttt{Simba}-\texttt{C} simulation's median results are shown by the blue line with its spread in the light blue band, while the red line displays the median \texttt{Simba} results for comparison.}
    \label{fig: msdfunc}
\end{figure}

In Fig. \ref{fig: msdfunc}, we show the comparison between the \texttt{Simba-C} and \texttt{Simba} simulations' dust mass functions at $z=0$. For comparison with observational results, we show results from \citet{Dunne2011}, who used data from Herschel Astrophysical Terahertz Large Area Survey (Herschel-ATLAS), and observational results from \citet{Clemens2013}, who used data from the Wide field Infrared Survey Explorer (WISE), the Spitzer space telescope, Infrared Astronomical Satellite (IRAS), and Herschel. The general trend shows that the dust mass function in the \texttt{Simba-C} simulation obtained lower values compared to the \texttt{Simba} simulation. However, it remains in line with the trend observed in \citet{Dunne2011}. From subsequent comparisons conducted, we can also confirm that there were no other discernible changes resulting from the inclusion of dust in the simulations. This is likely due to the dust sputtering back into metals in the hot diffused gas, therefore minimizing its effect in the group regime. However, this could be different for cold gas, which is not explored here. Due to this, the \texttt{Simba-C NoDust} simulation will not be shown in the subsequent comparison tests.

\subsection{Runs and analysis}

We use four different \texttt{Simba}/\texttt{Simba-C} simulations, each incorporating a different physics module, until we reach the complete \texttt{Simba-C} model as described in Sec. \ref{sec: Simba-C simulation}, with dust included. The complete \texttt{Simba-C} and \texttt{Simba} simulations (iii and iv) have volumes of side length $100\, \mathrm{Mpc} \, h^{-1}$, consisting of $1024^{3}$ gas particles and $1024^{3}$ dark-matter particles, while the other two simulations (i and ii) have volumes of side-length $50\, \mathrm{Mpc} \, h^{-1}$ with $512^{3}$ gas particles and $512^{3}$ dark-matter particles and are used only for reference. All simulations ran from an initial redshift of $z=249$, down to $z=0$ and follow a \citet{Planck2018} $\Lambda$CDM cosmology of $\Omega_{m}=0.3$, $\Omega_{\Lambda}=0.7$, $\Omega_{b} = 0.048$, and $H_{0} = 68 \mathrm{\, km \, s^{-1} \, Mpc^{-1}}$. The different versions of the simulations are as follows:
\begin{enumerate}[label=(\roman*), leftmargin=2.5\parindent, labelwidth=\parindent]
    \item The \texttt{Simba} simulation without any included feedback mechanisms, but metal injection from stellar evolution is still present -- \texttt{Simba NoFeedback}. 
    \item The \texttt{Simba} simulation utilizing the instantaneous recycling of the metals stellar feedback model, but excluding AGN feedback. This configuration resembles the simulation used in \citet{Liang2016} -- \texttt{Simba NoAGN}.
    \item The original \texttt{Simba} simulation, as described in \citet{Dave2019}, which contains the instantaneous metal recycling model, the updated AGN feedback model,  and the described dust model -- \texttt{Simba}.
    \item The complete \texttt{Simba-C} simulation incorporating the newly updated chemical enrichment and stellar feedback model, the recalibrated AGN feedback model from the original \texttt{Simba} simulation, and the re-integrated dust model. We consider this to be our main simulation/result of the study -- \texttt{Simba-C}.
\end{enumerate}
 
We analyze the simulation outputs using a friends-of-friends galaxy finder, assuming a spatial linking length of $0.0056 \times$ the mean interparticle spacing, as well as the AMIGA Halo Finder (\texttt{AHF}), a tool developed by \citet{Knebe2008} and \citet{Knollmann2009} specifically designed for halo identification. Post-processing involves cross-matching galaxies and haloes using two distinct Python packages: \texttt{Caesar} and \texttt{XIGrM}, each serving specific purposes in computing group properties.
\begin{enumerate}[label=(\roman*), leftmargin=2.5\parindent, labelwidth=\parindent]
    \item \texttt{Caesar}:\footnote{The Caesar documentation can be found at \url{https://caesar.readthedocs.io/en/latest/}.} This \texttt{yt}-based Python package performs galaxy finding which is applied to all stars, black holes, and cool gas elements above a specified minimum SF threshold density of $n_{H} > 0.13\,\mathrm{ H\, atoms\, cm}^{-3}$. Black holes are associated with galaxies to which they are most gravitationally bound, and the most massive black hole within a galaxy is designated as the central black hole. 
    \item \texttt{XIGrM}:\footnote{The XIGrM documentation can be found at \url{https://xigrm.readthedocs.io/en/latest/}.} This Python package specializes in computing the X-ray properties of the IntraGroup Medium, to compute the group properties as detailed below.
    
\end{enumerate}

Previous studies exploring the effects of AGN feedback has been done by comparing the \texttt{Simba NoAGN} simulation to the \texttt{Simba} simulation (see \citet{Robson2020, Robson2021} and \citet{Chowdhury2022}, and references therein). Despite these previous investigations, we include the \texttt{Simba NoAGN} simulation for reference, since we recalibrated the strength of the AGN feedback in \citet{Hough2023}. Discussions based on the impacted properties are given in Sec. \ref{sec: ST scaling}.

\subsection{Computing group properties}
\subsubsection{Finding the galaxy group haloes}

To identify galaxies and galaxy groups within each simulation, analysis is performed for outputs corresponding to redshifts $z=2,1,0.5$ and $z=0$. We follow the method used in \citet{Liang2016}, \citet{Jung2022}, and \citet{Saeedzadeh2023}. Similar studies using different cosmological simulations, such as the Feedback Acting on Baryons in Large-scale Environments (\texttt{FABLE}) simulation \citep{Helden2018} have also been done. We utilized the AHF software to find hierarchy structures nested for haloes and sub-haloes, by locating peaks in the adaptively smoothed density field, through the identification of all particles (gas, stars, dark matter, and black holes) that are gravitationally bound to each other \citep{Jung2022, Saeedzadeh2023}. We then proceed up in the hierarchy to find the larger structures, while the centres of these haloes are located by applying the shrinking-sphere approach \citep{Power2003}. The determination of halo masses ($\mathrm{M_{\Delta}}$) involves the construction of a sphere and the expansion of its radius until the total mass density (enclosed) equals the viral density for a specific cosmology at a certain redshift. 
\begin{equation}
    \rho_{m,\Delta}(z) = \Delta\cdot E^{2}(z)\rho_{\mathrm{crit}}(0),
\end{equation}
with $E(z) \equiv H(z)/H_{0}$ being the dimensionless Hubble parameter given by:
\begin{equation}
    E(z)^{2} = 1-\Omega_{\mathrm{m,0}}+\Omega_{\mathrm{m,0}}(1+z)^{3},
\end{equation}
and $\rho_{\mathrm{crit}}$ being the critical cosmology density.\footnote{The $R_{\Delta}$ of the group is defined
as the radius within which the mean density of the group is $\Delta\times \rho_{\mathrm{crit}}$, at the group's redshift.} We also use the frequently used $\Delta$ values, that is, 200, 500, and 2500 in addition to the virial radius/mass\footnote{The mapping between these quantities is redshift dependent. This dependency on the redshift is taken into account by using the $E(z)$ factor.} \citep{Babul2002, Lovisari2021a, Saeedzadeh2023}. The following conversion equations are also used: 
\begin{equation}
    \begin{split}
        M_{500} &= 500\times \frac{4}{3} \pi R_{500}^{3}\,\rho_{\mathrm{crit}},\\
        R_{2500} &\approx R_{500}\times 0.4.
    \end{split}
\end{equation}

We utilize the Python package XIGrM to determine the various halo quantities, including the radius, mass, luminosity function, entropy of the system, and the temperature of the host halo at each $\Delta$-value.  

\subsubsection{Group mass functions}
\begin{figure}
\includegraphics[width=1.1\columnwidth, trim=0.8cm 0.6cm 0cm 1.1cm, clip]{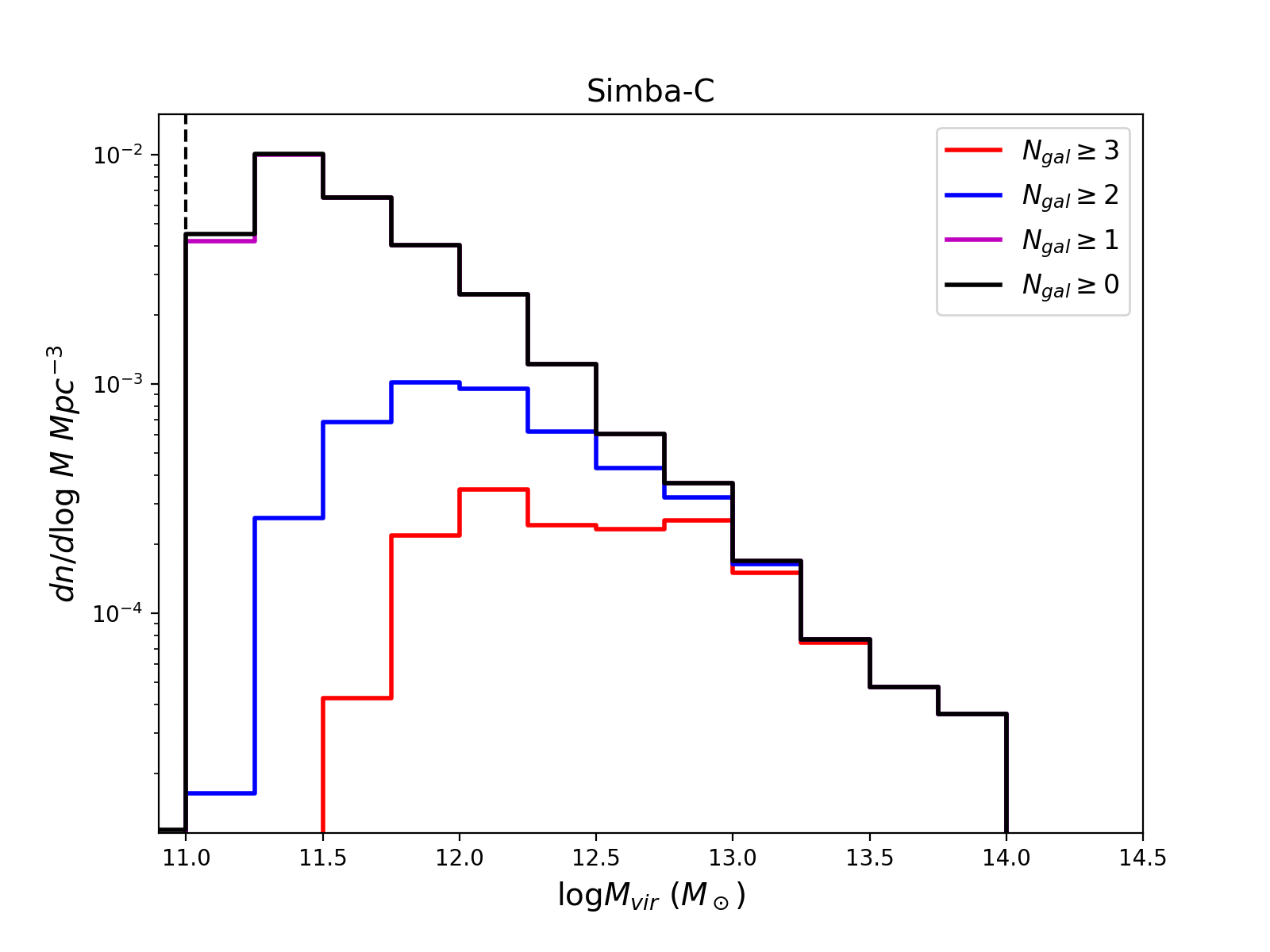}  
\caption{The halo mass function for haloes with at least three (red), two (blue), or one (magenta) luminous galaxies, as well as the complete halo population in the simulation (black). The dashed vertical line shows the halo mass ($10^{11}\,\mathrm{ M_{\odot}}$) cut-off introduced in the XIGrM halo analysis. All \texttt{Simba}-based simulations have a halo mass resolution of $6.8\times 10^{9}\,\mathrm{ M_{\odot}}$, corresponding to 64 dark matter particles.}
    \label{fig: Halo MF z=0}
\end{figure}

In Fig. \ref{fig: Halo MF z=0}, we present the $z=0$ halo mass function for the full \texttt{Simba-C} simulation (the halo mass functions of the other three simulations are very similar). The black curve represents all haloes, whereas the red, blue, and magenta curves depict haloes with at least three, two, and one `luminous' galaxies, respectively. A `luminous' galaxy is defined as having a stellar mass of at least $M_{*} \sim 1.16\times 10^{9}\, \mathrm{ M_{\odot}}$, equivalent to at least 64 star particles, with each particle having a mass resolution of approximately $\sim 1.8\times 10^{7}\,\mathrm{M_{\odot}}$. Per definition, groups and clusters are identified as haloes with three or more luminous galaxies \citep{Liang2016}. Consistent with \citet{Liang2016}, we observed that on mass scales of $\geq 10^{13}\, \mathrm{M_{\odot}}$, nearly all haloes have at least three luminous galaxies.

\begin{figure}
\includegraphics[width=1.035\columnwidth, trim=0.2cm 0.2cm 0cm 0.3cm, clip]{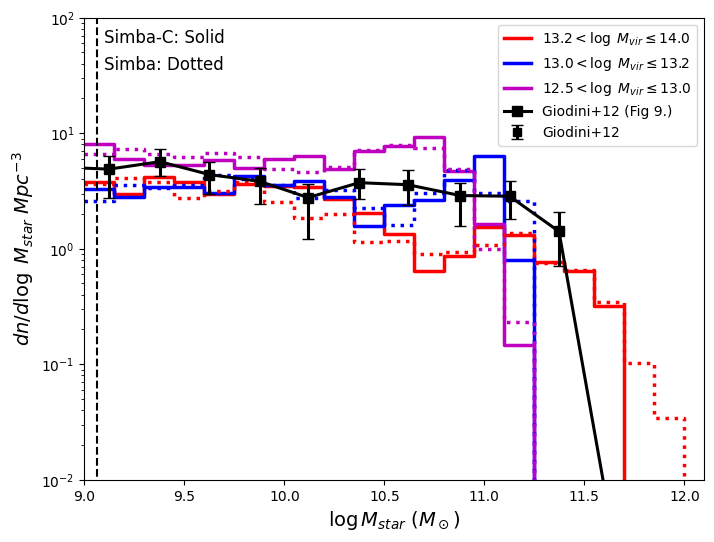}
    \caption{The galaxy stellar mass function (GSMF) for all luminous galaxies in the simulated groups, sorted into three mass bins: $12.5< \log M_{vir}\leq 13.0\,\mathrm{ M_{\odot}}$ (magenta), $13.0< \log M_{vir}\leq 13.2\,\mathrm{ M_{\odot}}$ (blue), $13.4< \log M_{vir}\leq 14.0\,\mathrm{ M_{\odot}}$ (red). Simulated galaxies from \texttt{Simba-C} is shown with solid lines, while \texttt{Simba} is shown with dotted lines. We compare this to observations of the GSMF for low-mass X-ray detected groups that span the same mass range as our simulated groups \citep{Giodini2012}. We artificially reduced the galaxy's stellar mass function in the simulation by a factor of three, as described in the text. The vertically dashed black line shows our luminous galaxy's stellar mass resolution of $M_{*} \sim 1.16\times 10^{9}\, \mathrm{ M_{\odot}}$.}
    \label{fig: GSMF z=0}
\end{figure}

In Fig. \ref{fig: GSMF z=0}, the average galaxy stellar mass function (GSMF) for all luminous galaxies in their host groups are shown for the complete \texttt{Simba-C} (solid lines) and the original \texttt{Simba} simulations (dotted lines). The groups are categorized into mass bins corresponding to $12.5 < \log M_{\mathrm{vir}} \leq 13.0\,\mathrm{ M_{\odot}}$ (magenta), $13.0 < \log M_{\mathrm{vir}} \leq 13.2\,\mathrm{ M_{\odot}}$ (blue), and $13.4 < \log M_{\mathrm{vir}} \leq 14.0\,\mathrm{ M_{\odot}}$ (red). These mass bins align with those used in \citet{Dave2008} and \citet{Liang2016}. In addition, a comparison is made with the observed GSMF for low-mass X-ray detected groups in the Cosmic Evolution Survey (COSMOS) of \citet{Giodini2012}. It is important to note that, for comparison purposes, the galaxy's stellar mass function in all three mass ranges has been artificially reduced by a factor of three for both simulations. This adjustment is made because of the unknown volume of the galaxy groups within the catalogues relative to the observations, which is treated as a free parameter. This artificially reduced value has no physical meaning. Hence, the only predictive power for comparison with the data is in the shape of the group stellar mass function.

The shape of the binned galaxy stellar mass functions for both simulations is shown in Fig. \ref{fig: GSMF z=0}, matches the overall observational trend fairly well, particularly for the highest mass bin that is closest to the observed sample. Notably, unlike in \citet{Liang2016}, there is no excess of galaxies with very large stellar masses in \texttt{Simba-C}. \texttt{Simba} does have an excess, although much smaller than in \citet{Liang2016}, which did not include AGN feedback. \citet{Liang2016} suggested that the introduction of AGN feedback would quench the `supersized' central galaxies. This assertion is confirmed when using the \texttt{Simba NoAGN} simulation, which did indeed show an excess of `supersized' central galaxies. Therefore, similarly to the overall galaxy mass function, the group GSMF also requires AGN feedback to be accurately reproduced in models.

\subsubsection{Global X-ray galaxy group properties} \label{sec: Xray properties}
Once we identify the haloes, the next step is to compute the various observed X-ray properties. To simplify our analysis, we focus on the hot diffused IGrM gas particles.  These are defined as gas particles with temperatures above the threshold of $T>5\times 10^{5}\,\mathrm{ K}$ with a hydrogen number density ($n_{h} < 0.13\, \mathrm{ H\, atoms\, cm}^{-3}$), which is below the density threshold to allow star formation. 

The first X-ray property computed is the rest-frame $0.5-2.0\, \mathrm{ keV}$ luminosity within $R_{500}$, denoted as $L_{X,0.5-2.0}$. To calculate this property, the luminosity of each individual IGrM gas particle within a distance $r \leq R_{500}$ from the halo centre is summed. The emission characteristics of the gas particles are computed using the Python package \texttt{PyAtomDB},\footnote{The PyAtomDB documentation can be found at \url{https://atomdb.readthedocs.io/en/master/}.} which in turn uses the Atomic Data for Astrophysicist database (AtomDB), which itself consists of two components: (i) Astrophysical Plasma Emission Database (APED) and (ii) Astrophysical Plasma Emission Code (APEC; \citealt{Smith2001}). The gas is assumed to be optically thin and to be in collisional ionisation equilibrium. APEC uses the particle's mass, SPH-weighted density, temperature, and metallicity as input while returning an X-ray spectrum. By summing the photon energy intensities in the specified energy range, the luminosity is obtained. Contributions to line and continuum emissions associated with each tracked element in \texttt{Simba-C} are computed separately, except for Gallium and Germanium due to the 30-element limit of AtomDB. 

We use mass-weighted (mw) and emission-weighted (ew) abundances to study the metal content of the IGrM. They are defined as
\begin{equation}
    Z_{q}^{mw} = \frac{\sum_{i}Z_{q,i}m_{i}}{\sum_{i}^{N}m_{i}}\; \mathrm{and} \; 
    Z_{q}^{ew} =\frac{\sum_{i}Z_{q,i}L_{i}}{\sum_{i}^{N}L_{i}},
\end{equation}
where $q$ is the metal species, $Z_{q,i}$ is the SPH kernel-weighted abundance of the $i^{th}$ particle, $m_{i}$ and $L_{i}$ its mass and X-ray luminosity, respectively. The sum runs over all IGrM particles within the halo's volume. Furthermore, all metal abundance estimates are in terms of solar `photosphere abundance' values from \citet{Anders1989}.\footnote{This is different than the relative `photosphere abundances' values that \texttt{Simba} and \texttt{Simba-C} uses from \citet{Asplund2009}.}

In the calculation of X-ray properties, the spectroscopic temperature $T_{spec}$ is used rather than the frequently used $T_{X}$ temperature, where $T_{X}$ represents an average of the temperatures of the individual components weighted by their contributions to radiative emission. The choice of $T_{spec}$ over $T_{X}$ is motivated by the tendency of $T_{X}$ to be biased high by approximately $25\%$ for clusters \citep{Mazzotta2004}. In contrast, $T_{spec}$ is designed as a weighting scheme to produce a temperature that is comparable to the temperatures of a hot gas determined by observations of groups and clusters \citep{Vikhlinin2006}. Specifically, $T_{spec}$ is determined by identifying a single-temperature thermal model whose spectrum best matches the observed spectrum.
 
In the X-ray analysis, the decoupled hot-wind particles, representing a fraction of the stellar feedback-driven wind particles that are heated, are excluded. The fraction of these particles in \texttt{Simba-C} follows the trends of the FIRE simulations \citep{Pandya2022}, while in \texttt{Simba}, it was a fixed fraction of 30\%. Although these particles constitute a very small portion of the total mass \citep{Appleby2021}, due to their high density as they emerge from the ISM, they have extreme X-ray luminosities. Two different X-ray calculations are performed: (i) All the hot diffuse gas particles within $R_{500}$ are used, or (ii) only the hot diffuse gas particles within the radial range $0.15R_{500}\leq r \leq R_{500}$ are used. The latter is referred to as `core-excised' \citep[e.g.,][]{Liang2016}, which is often more robust, as the cores of clusters and groups can have widely varying temperatures. Hence, observations often present `core-excised' temperatures to which we will compare where appropriate.

We first study direct comparisons to observations at present-day ($z \approx 0$) for each of our four simulations, to obtain an understanding of which physics models/properties play the largest role in improving the simulation for each scaling relation in Sec. \ref{sec: scaling relations}. The expected largest impact physics model is the addition of AGN as shown in \citet{Robson2020}; however, other effects may also have a significant contribution, depending on the tested property; therefore, all combinations must be tested. Then, using this knowledge, we look at how each simulation evolves with redshift in Sec. \ref{sec: Redshift evolution}. This will give us the necessary insight to understand how the introduction of each physics module played its role in improving the evolution of the galaxy groups within simulations. Lastly, given the proven successes of the \texttt{Simba} simulation with certain X-ray properties, e.g. the mass fractions in \citet{Robson2020}, we also show the trends of the \texttt{Simba} simulation to confirm that the improvements made in \texttt{Simba-C} did not adversely affect these predictions and to identify any improvements that \texttt{Simba-C} has compared to its predecessor.

We follow the convention in the literature and plot the quantities motivated by the self-similar model for group and cluster haloes \citep{Kaiser1986}. In this model, the scaling relations are preserved when using the quantities: (i) $L_{X}(z)E(z)^{-1}$, (ii) $M_{\Delta}(z)E(z)$, and (iii) $S_{\Delta}(z)E(z)^{4/3}$, with $E(z) \equiv H(z)/H_{0}$ being the dimensionless Hubble parameter.

\section{Global scaling relations and halo structure comparisons to present-day observations}\label{sec: scaling relations}

We begin by presenting comparisons to observations of the X-ray scaling relations and baryon mass fractions at $z=0$ for each of our four simulations. This comparison will help assess the capability of the \texttt{Simba-C} simulation to capture the underlying physics that drives the X-ray properties of galaxy groups. In this section, the spectroscopic weighted temperature is used, as discussed in Section \ref{sec: Xray properties}.

\subsection{Luminosity-Temperature scaling relation}\label{sec: LT scaling}

We first consider the scaling relation between the X-ray luminosity $L_X$ and the X-ray spectroscopic temperature $T_{spec}$.  Observations generally indicate a steeper scaling relation ($L_{X}\propto T^{3-5}$) for the lowest mass groups, whereas massive clusters generally align with the predicted slope of the self-similar model~\citep{Balogh1999, Robson2021, Lovisari2021a}, that is, $L_{X}\propto T^{1-2}$,\footnote{This slope in it self is also dependent on the whether working with relaxed clusters or note. For example, \citet{Pratt2009} measured a $L_{X}\propto T^{2.7-2.9}$ slope.} depending on the X-ray passband under consideration. Two primary physical effects impact this scaling relation. First, there is the radiation mechanism. In clusters ($T_X\ga 1$~keV), the self-similar model predicts the bolometric luminosity scaling as $L\sim T^{2}$ due to the bremsstrahlung being the dominant mechanism. In groups, line radiation begins to dominate emission. However, this predicts a \emph{flatter, not a steeper} $L-T$ relationship \citep{Balogh1999}. Second, feedback affects the IGrM, influencing both the amount of gas within the IGrM and its radial profile. The overall impact is more pronounced in the lowest-mass systems due to their smaller gas reservoirs and potential wells. This significantly influences $L_X$ due to its density-square dependence. For example, \citet{Balogh1999} model groups with isentropic cores show a $L\sim T^{5}$ slope, as observed in low-mass groups. According to \citet{Balogh1999, Voit2001, Babul2002, McCarthy2004}, this indicates that heating and/or cooling have substantially altered the distribution of the hot X-ray gas. Hence, by probing scaling relations, one can obtain constraints on the group's feedback by accounting for the line emission mechanism.

\begin{figure}
\includegraphics[width=\columnwidth, trim=0.2cm 0.2cm 0.3cm 0.2cm,clip]{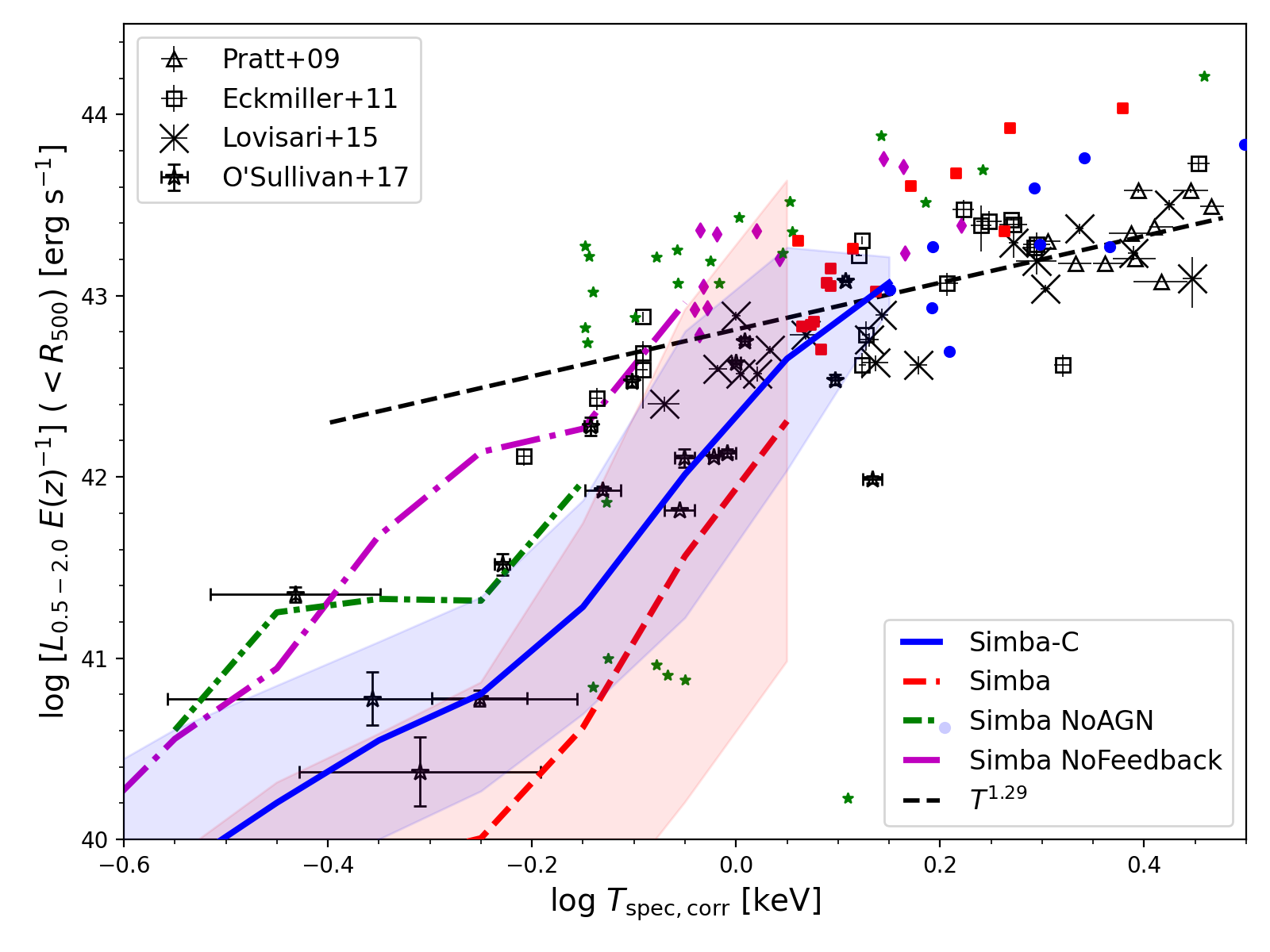}
    \caption{$L_{X,0.5-2.0}-T_{spec,corr}$ relation within $R_{500}$ for the simulated groups in the $0.5-2.0\,\mathrm{ keV}$ band for the various simulations: \texttt{Simba-C} (blue circle/solid line), \texttt{Simba} (red square/dashed line), \texttt{Simba NoAGN} (green star/tight dot-dashed line), and \texttt{Simba NoFeedback} (magenta diamond/loosely dot-dashed line). The light blue and red shaded bands show the $1\sigma$-deviation for the \texttt{Simba-C} and \texttt{Simba} simulated haloes, respectively. For comparison, observations of the following low-redshift group data are included: \citet{Pratt2009} (triangles), \citet{Eckmiller2011} (squares), \citet{Lovisari2015} (crosses), and \citet{OSullivan2017} (stars). The dashed black line shows the theoretical predictions based on the self-similar model's power-law ($L \propto T^{1.29}$) over the temperature range $0.4-3\,\mathrm{keV}$ from Table 1 in \citet{Lovisari2021a}.}
    \label{fig: LT_spec_corr}
\end{figure}

Fig. \ref{fig: LT_spec_corr} shows the rest frame $0.5-2.0\,\mathrm{ keV}$ X-ray luminosity emitted within $R_{500}$ against the mean core-excised spectroscopic temperature for the simulated groups in our four simulations at $z=0$: (i) The complete \texttt{Simba-C} simulation (blue solid line), (ii) the original \texttt{Simba} simulation (red dashed line), (iii) the \texttt{Simba NoAGN} simulation (green tight dot-dashed line), and finally (iv) the \texttt{Simba NoFeedback} (magenta loosely dot-dashed line). For our two main comparison simulation (\texttt{Simba-C/Simba}) results, we show a light blue and a light red band to display the $1\sigma$-deviation in each temperature bin for each simulation, respectively. We include at least 10 haloes in each temperature bin (see Table \ref{tab: halo amounts in each temp bin}). In cases where a temperature bin contains less than 10 haloes, the individual halo values are not binned and are represented with the following markers: (i) \texttt{Simba-C} (blue circles), (ii) \texttt{Simba} (red squares), (iii) \texttt{Simba NoAGN} (green stars), and (iv) \texttt{Simba NoFeedback} (magenta diamonds). This approach provides some insight into the emerging trends for the more massive galaxy groups and clusters, even where the trends cannot necessarily be confirmed as statistically significant because there are fewer than 10 haloes per bin. This approach will be consistently applied throughout Sec. \ref{sec: scaling relations}.

\begin{table*}    
\centering
\begin{tabular}{|c|c|c|c|c|c|c|c|c|c|c|c|c|c|}
\hline
Simulation &$T_{spec,corr} = -0.7$ & -0.6 &-0.5 &-0.4 &-0.3 &-0.2 &-0.1  &0  &0.1  &0.2  &0.3 & 0.4\\
\hline
\texttt{Simba-C} ($100\, \mathrm{Mpc} \, h^{-1}$)& 36  &80 &191 &314 &303 &204 &107  &46  &16  &13   &2   &3  \\
\texttt{Simba} ($100\, \mathrm{Mpc} \, h^{-1}$)& 58  &84 &185 &218 &240 &183 &119  &46  &27   &4   &4   &1  \\
\texttt{Simba NoAGN feedback} ($50\, \mathrm{Mpc} \, h^{-1}$)& 8  &7 &20 &44 &73 &55 &23  &9  &4  &3  &1  &0 \\
\texttt{Simba NoFeedback} ($50\, \mathrm{Mpc} \, h^{-1}$)& 115 &166 &168  &84  &34  &20  &15  &14   &2   &3   &1   &0  \\
\hline
\end{tabular}
\caption{The number of haloes in each temperature bin is provided for each simulation across all $T_{spec,corr}$ plots. The $T_{spec,corr}$ value shown in the bin names denotes the lower $T_{spec,corr}$ value of the bin. All bins increase with a $\log T_{spec,corr}=0.1$ bin size.}
\label{tab: halo amounts in each temp bin}
\end{table*}

For comparison, we have included low redshift X-ray observations from \citet{Pratt2009}, using data from the Representative \textit{XMM-Newton} Cluster Structure Survey (REXCESS), \citet{Eckmiller2011} using the Highest X-ray FLUx Galaxy Cluster Sample (HIFLUGCS), \citet{Lovisari2015} using \textit{XMM-Newton} observations for a complete sample of galaxy groups, and finally \citet{OSullivan2017} using the Complete Local Volume Groups Sample (CLoGS) from \textit{XMM-Newton} and \textit{Chandra} observations.\footnote{Note that \citet{OSullivan2017} provide X-ray luminosities in the $0.7-5.0\,\mathrm{keV}$ band. We applied a temperature-dependent correction factor to estimate the corresponding $0.5-2.0\,\mathrm{keV}$ luminosity.} In addition, we include a black dashed line representing the `self-similar' model results in the $0.5-2.0\,\mathrm{keV}$ band, incorporating both line radiation and bremsstrahlung. This self-similar model adopts the slopes from Table 1 in \citet{Lovisari2021a}.

From Fig. \ref{fig: LT_spec_corr}, we note that the complete \texttt{Simba-C} simulation and its $1\sigma$-error range generally overlap with the observations, although very slightly low for the coldest groups. Following the trends of individual halo values for the most massive groups and clusters, the \texttt{Simba-C} simulations seem to match the observations the most closely. This suggests that \texttt{Simba-C} broadly succeeds in determining $L_X$ through a combination of cooling and feedback. Compared to \texttt{Simba-C}, \texttt{Simba} has reduced luminosities in the lower halo mass regime, to the point where \texttt{Simba} obtained too low luminosities compared to observations, although some observations are still in its $1\sigma$-error. This improvement from \texttt{Simba-C} likely stems from the overall lower metal mass fractions in \texttt{Simba-C} when adopting the new chemical evolution model, as demonstrated in Sec. \ref{sec: Metal enrichment of IGrM}, which reduces metal-line cooling and consequently yields more hot gas. In the higher halo-mass regime, the opposite trend is observed, with individual haloes tending to be slightly too bright on average. However, when taking into account the $1\sigma$-deviations, the improvements obtained by \texttt{Simba-C} over \texttt{Simba} in the $L_{X,0.5-2.0}-T_{spec,corr}$ relation are only significant in the colder groups, while in the warm groups these differences are insignificant.

Furthermore, Fig. \ref{fig: LT_spec_corr} shows that the `self-similar' model tends to overestimate the $L_{X,0.5-2.0}-T_{spec,corr}$ relation compared to observations.  This well-known result is commonly attributed to feedback and/or cooling, which selectively removes hot gas more from smaller systems. Among our simulations, the NoFeedback run is the most similar to the self-similar model. Therefore, the impact of cooling can be assessed from this model and it alone is not sufficient to explain observations (see also \citealt{McCarthy2004, McCarthy2008}).  The NoAGN model has a lower $L_X$ value than the NoFeedback model, but still generally exceeds the observations and is still higher than the two main simulation results, suggesting that stellar feedback plays a role in gas removal (see also \citealt{Liang2016}), but that AGN feedback remains the primary $L_{X}$ reduction source \citet{Robson2020}.

\subsection{Mass-Temperature scaling relation}\label{sec: MT scaling}

The relationship between mass and temperature is not anticipated to be very sensitive to baryonic processes, as for bound systems, the temperature should predominantly reflect the gravitational potential primarily driven by dark matter.  However, modest departures from self-similarity can still occur due to the interaction between feedback heating and gas removal in lower-mass systems, impacting $T_{spec}$. 

\begin{figure}
\includegraphics[width=\columnwidth, trim=0.2cm 0.2cm 0.2cm 0.2cm,clip]{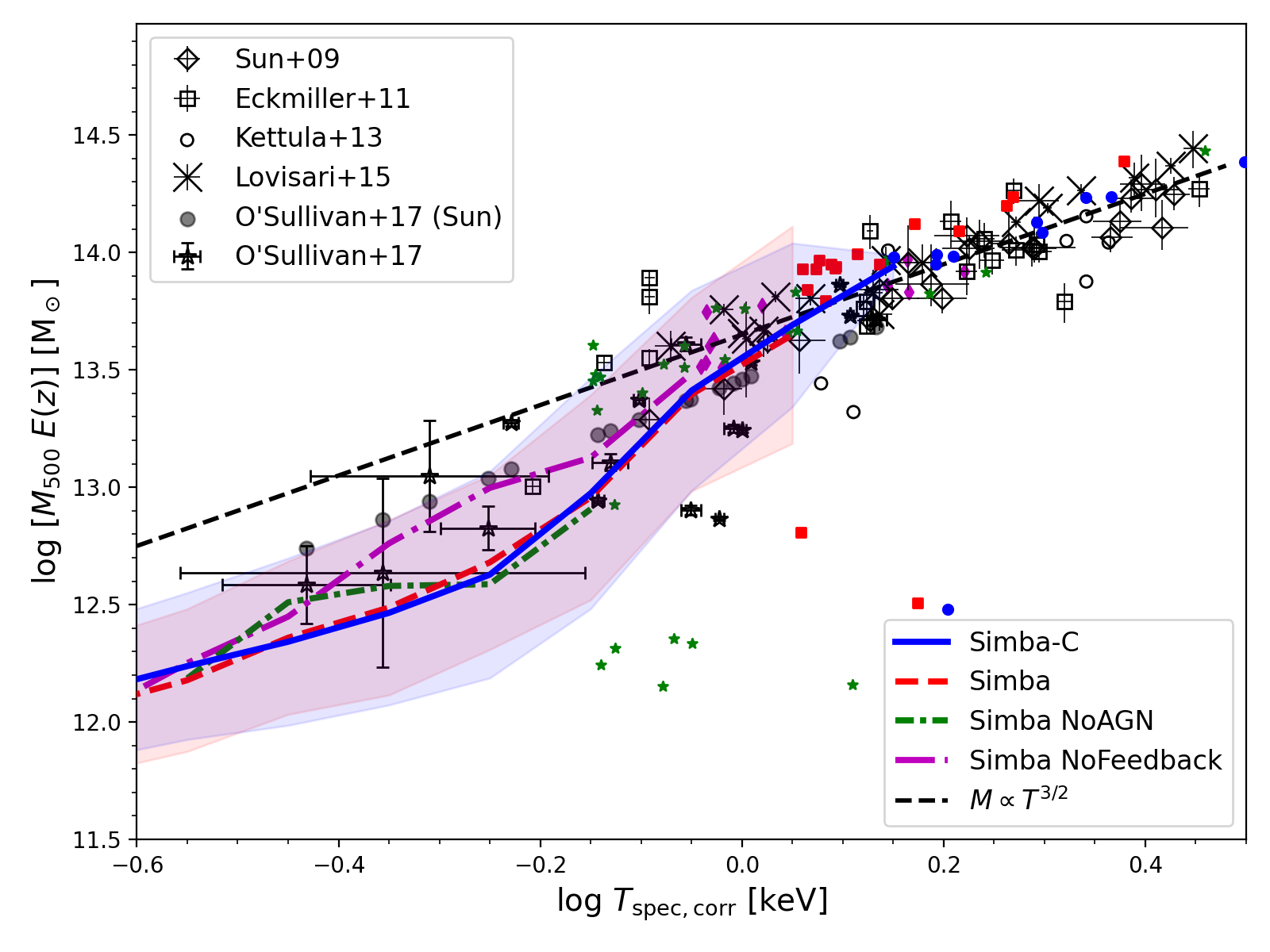}
    \caption{$M_{500}-T_{spec,corr}$ relation for the various different simulations with their lines as described in Fig. \ref{fig: LT_spec_corr}. For comparison, observations of the following low redshift group data are included: \citet{Sun2009} (diamonds), \citet{Eckmiller2011} (squares), \citet{Kettula2013} (circles), \citet{Lovisari2015} (crosses), and \citet{OSullivan2017} (grey-filled circles - \citet{Sun2009} estimate). The revised mass estimates for \citet{OSullivan2017} (stars), are based on \texttt{Simba-C}'s $M_{500}-T_{spec,corr}$ power-law function. The dashed black line shows the self-similar model ($M\propto T^{1.5}$) from Table 2 in \citet{Lovisari2021a}. The hydrostatic mass estimates from \citet{Eckmiller2011} and \citet{Sun2009} have been corrected for the hydrostatic bias \citep{Hoekstra2015}.}
    \label{fig: MT_spec_corr}
\end{figure}

Fig. \ref{fig: MT_spec_corr} shows the $M-T_{spec,corr}$ relation for the mass of the simulated groups within the central $R_{500}$ region for our four simulations at $z=0$. For comparison, we included low-redshift X-ray observational results from \citet{Sun2009} using \textit{Chandra} archival data, \citet{Eckmiller2011} using the HIFLUGC Survey, \citet{Kettula2013} using COSMOS results, and \citet{Lovisari2015} with their \textit{XMM-Newton} observations. We also present the original \citet{OSullivan2017} CLoGS results based on the $M-T_{spec,corr}$ relation of \citet{Sun2009}, as well as the revised CLoGS results utilizing \texttt{SIMBA-C}'s $M_{500}-T_{spec,corr}$ relationship for $-0.6\leq \log (T_{spec})\leq 0$ to determine the mass of the CLoGS groups.\footnote{Previously, the CLoGS masses were estimated using the scaling relations of Tier 1+2 groups from \citet{Sun2009}.}  Consequently, we cannot draw conclusions about our simulations that match the CLoGS results, but it provides insight into the observational trends of the lower-temperature groups. We use our updated results when examining the IGrM entropy in Fig. \ref{fig: ST_spec_corr}.

As expected, Fig. \ref{fig: MT_spec_corr} reveals no strong differences between simulations for this particular scaling relation. The four simulations produced $M_{500}-T_{spec,corr}$ curves, which are within $1\sigma$ of each other in all temperature ranges. The scaling is also consistent with that obtained by \citet{Liang2016} using a simulation without AGN feedback. They all demonstrate reduced masses for a given $T_{spec}$ on group scales, indicating departures from self-similarity driven primarily by the interplay of radiative cooling (present in all models) and the measurement of $T_{spec}$. Furthermore, we can also conclude that $T_{spec,corr}$ can be considered as a proxy for the halo mass. This property allowed us to retroactively determine the CLoGS's $M_{500}$ values used here. However, temperature is consistently used in further plots due to its direct comparison with observations, as used in the CLoGS paper \citep{OSullivan2017}, and other galaxy group X-ray scaling relation property studies, e.g., \citet{Lovisari2021a}. 

All simulations closely follow the observational trends, even though most of the overlapping observational results are from the CLoGS sample, which is to an extent by construction. This holds true even for the NoFeedback run, since this scaling relation is more determined by gravitational processes than by feedback interactions; i.e., the gas expands until its temperature is consistent with the gravitational potential of the group. Hence, while this scaling relation does not provide discriminatory power between models, it is reassuring that the simulations can reproduce the observations down to the smallest systems and support the viability of \texttt{Simba-C} for X-ray group studies.

\subsection{Entropy-Temperature scaling}\label{sec: ST scaling}

\begin{figure}
\includegraphics[width=\columnwidth, trim=0.5cm 0.5cm 0.3cm 0.1cm,clip]{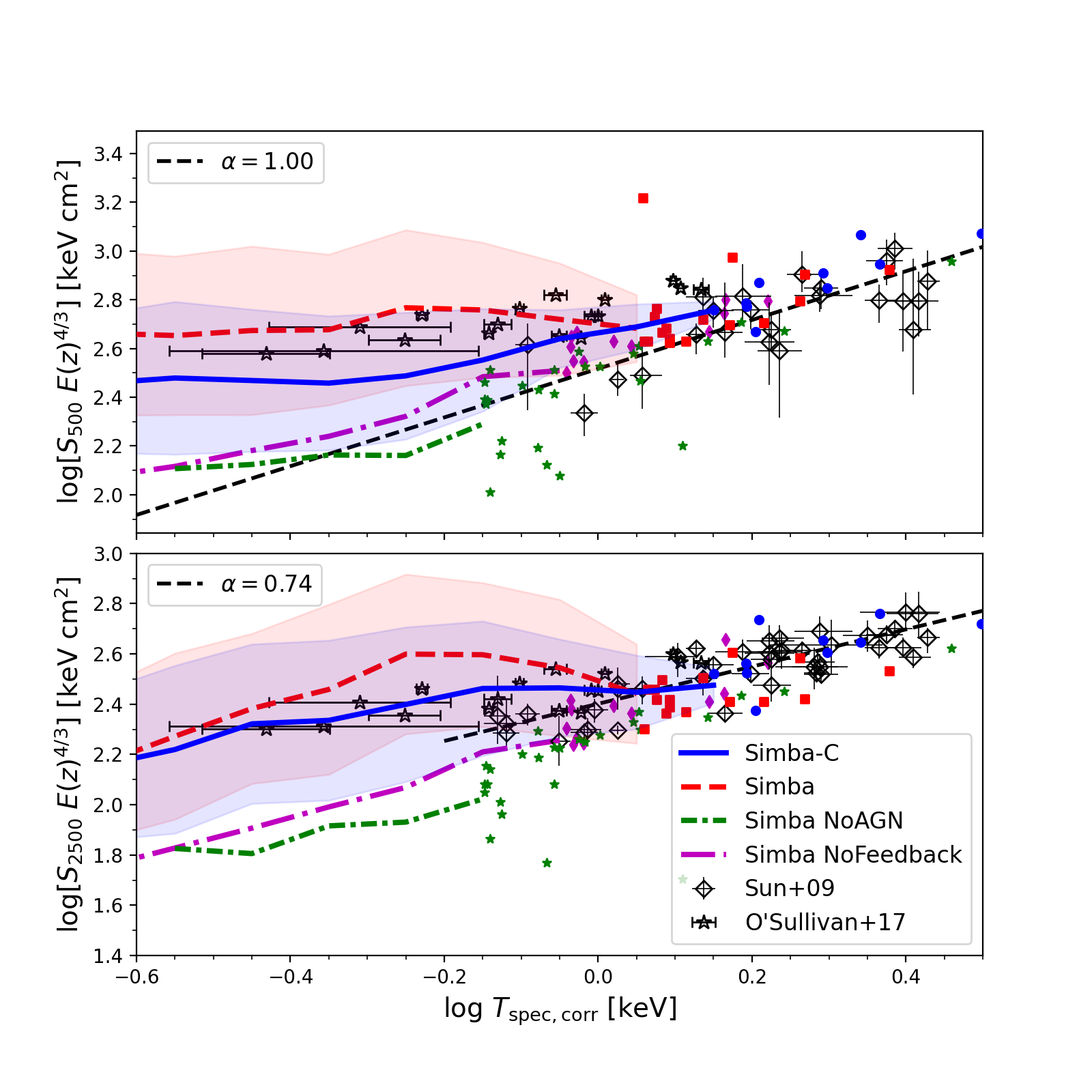}
    \caption{Gas entropy at $R_{500}$ and $R_{2500}$ for the various simulations with their lines as described in Fig. \ref{fig: LT_spec_corr}. Observations of the following low redshift group data are included for comparison: \citet{Sun2009} (diamonds) and \citet{OSullivan2017} (stars). The dashed lines are the best-fit power-law indices of $\alpha=1$ and $\alpha=0.74$ for the S-T relation at $R_{500}$ and $R_{2500}$, respectively, for the full group+cluster sample from \citet{Sun2009}.} %The entropy estimates were determined directly from figure 4 in \citet{OSullivan2017}, based on the power-law $S_{\Delta} \approx R_{\Delta}^{0.701}$, with $R_{\Delta}$ calculated from the estimated mass based on $T_{spec,corr}$.
    \label{fig: ST_spec_corr}
\end{figure}

As highlighted in \citet{Balogh1999}, \citet{Babul2002}, \citet{Lewis2000}, and \citet{Voit2001}, entropy serves as a valuable quantity to study how the IGrM is influenced by cooling and/or heating processes. This is because a significant portion of the Universe's baryons reside in intergalactic space and experience heating through gravitationally driven shocks \citep{Dave2001}. Once heated, they settle into gravitational potential wells and adopt the characteristic temperature of the enclosing dark matter. However, the mean intensity of the X-ray emissions from the baryons reflects the amount of non-gravitational energy. The emissivity of baryons depends on how severely they are compressed and how this injection affects the baryon distribution \citep{Balogh1999, Voit2001, McCarthy2004, McCarthy2008}. Stellar and AGN feedback can restrict this compression \citep{Babul2002, McCarthy2004}, thus reducing the X-ray luminosity. These processes are essential because gravitational-only processes would excessively produce the $0.5-2.0\,\mathrm{keV}$ X-ray background when establishing the entropy distribution \citep{Pen1999, Wu2000}. In other words, the lowest entropy (most compressible) gas needs to be eliminated \citep{Voit2001b}. Non-gravitational heating and radiative cooling/subsequent condensation are potential physical reasons for this change in the low entropy gas \citep{McCarthy2004}. Furthermore, a lower limit to the entropy of the intracluster increases the $L_{X}-T$ relation, because shallower potential wells of low temperature groups/clusters are less capable of overcoming resistance to compression \citep{Balogh1999, Voit2001, Babul2002}. We plot the canonical proxy for entropy\footnote{Refer to \S 3.2.1 of \citet{Balogh1999} for a discussion of the relationship between the two.} in Fig. \ref{fig: ST_spec_corr}, rather than the thermodynamic specific entropy, given by: 
\begin{equation}
    S(r) = \frac{k_{\mathrm{B}}T_{spec}(r)}{n_{e}(r)^{2/3}},
\end{equation}
where $k_{\mathrm{B}}$ is the Boltzmann constant and $n_{e}(r)$ is the electron number density within a thin spherical shell at radius $r$. The gas distribution is organized so that the lowest entropy is in the centre of the halo, while the highest entropy is in the outer limits of the group \citep{McCarthy2004, McCarthy2008}. For this scaling relation, we also show $S_{2500}$ since the most notable improvement can be seen in this regime.

In Fig. \ref{fig: ST_spec_corr}, we show the gas entropy versus the spectroscopic temperature measured at $R_{500}$ and $R_{2500}$ (the inner core of the halo) for the simulated groups in our four simulations at $z=0$. The entropy was calculated by taking the average on a radial shell between $R_{x}$ and $1.05\times R_{x}$. For comparison, we show low-redshift X-ray observational results from \citet{Sun2009}, using the \textit{Chandra} archival data, as well as the CLoGS results from \citet{OSullivan2017}, using \textit{XMM-Newton} and \textit{Chandra} observations. Lastly, we show with the dashed black lines, the power-law indices of $\alpha=1$ and $\alpha=0.74$ corresponding to the best-fit $R_{500}$ and $R_{2500}$ values as determined by \citet{Sun2009} for the scaling of the group entropy.

From Fig. \ref{fig: ST_spec_corr}, it is evident that the different physical processes (feedback, dust, enrichment, etc.) impact the entropy at both radii as expected. For \texttt{Simba/Simba-C}, which include both feedback mechanisms that severely inhibit gas compression, a higher entropy is measured. The increased entropy suppresses gas cooling and condensation further until there is no more gas below the cooling threshold to form stars. If feedback is inefficient, condensation will remove this gas from the intracluster medium. If the feedback is highly efficient, it will increase entropy and convect to the outer regions of the cluster, resulting in a decrease in the gas density \citep{Voit2001}. In return, this reduces the luminosity and steepens the $L_{X}-T$ relation (cf. \citealt{Balogh1999, McCarthy2004}) also seen in Fig. \ref{fig: LT_spec_corr}. 

When comparing these two simulations with the observations, we note that although the $1\sigma$-error between \texttt{Simba-C} and \texttt{Simba} overlaps with each other at both radii (especially in the inner-core region), the average gas entropy in the haloes tends to be lower in \texttt{Simba-C} than in \texttt{Simba}. This results in the entire $1\sigma$-error range of \texttt{Simba-C} also being lowered. This positively impacts the \texttt{Simba-C} simulation which now closely matches, on average, with the $S_{2500}$ gas entropy CLoGS observations \citep{OSullivan2017}. The impact on the $S_{500}$ results are less pronounced, with the CLoGS observations tending to be, on average, between the \texttt{Simba} and \texttt{Simba-C} results, but still within the $1\sigma$-error. 
Furthermore, by examining the individual halo entropies, the general trend for \texttt{Simba-C} simulation most closely follows the higher mass groups and the slope of the self-similar model. Although true at both radii, it is more notable in the inner-core region, with the \texttt{Simba} simulation's $S_{2500}$ tending to be a bit more flat. From this we can conclude that \texttt{Simba-C} improves the gas entropy in the group inner-cores with respect to \texttt{Simba}, even though their overlapping $1\sigma$-error indicates that this improvement is insignificant. This is an important finding, since \texttt{Simba}'s entropy profiles have been shown to flatten in an extended entropy core \citep{Oppenheimer2021, Altamura2023}. Therefore, with \texttt{Simba-C} obtaining, on average, a lower entropy compared to \texttt{Simba} at both radii, it goes in the direction of relieving some of this tension and requires further investigation of the entropy profiles to provide context for this improvement. This is being done in a follow-up study by \textit{Padawer-Blatt et al.} (in prep).

Furthermore, Fig. \ref{fig: ST_spec_corr} also shows the impact of the inefficient metal cooling that \texttt{Simba-C} obtained. In \citet{Hough2023} \S 2.6, it was found that with the introduction of the new stellar feedback and metals model (\texttt{Chem5}), the \texttt{Simba-C} simulation was underproducing metals. This led to a shallow mass-metallicity relation (MZR). The \texttt{Chem5} model inherently produced fewer metals, resulting in less efficient metal cooling. To address this problem, the strength of the AGN feedback (a heating process) in \texttt{Simba-C} was reduced.\footnote{For a detailed discussion of the relationship between metal cooling and AGN feedback, we refer the reader to the second-half of \S 2.1 of \citet{Jung2022}.} This reduction is now reflected in the slightly reduced gas entropy obtained from \texttt{Simba-C}. This decrease in entropy allowed the \texttt{Simba-C} simulation to partially resolve some of the overcorrection/overestimation of the gas entropy observed after the AGN feedback was included into the \texttt{Simba} simulation, i.e., going from \texttt{Simba NoAGN} to \texttt{Simba}, which increased the entropy by approximately $0.3\,\mathrm{dex}$). 

Simulations lacking feedback, especially AGN feedback, dominated by gravitational-only shocks, exhibit lower entropy values at $R_{2500}$ than the other simulations. This, in turn, leads to an overproduction of the X-ray luminosity (see Fig. \ref{fig: LT_spec_corr} and \citealt{McCarthy2004}). It should also be noted that the NoFeedback model (cooling only) is closely aligned with the self-similar model; however, this alignment does not have physical significance. When gas cooling is allowed, it initially becomes more centrally concentrated, and the system's luminosity increases \citep{McCarthy2004}. However, as the gas continues to cool, stars start to form, reducing the hot gas fraction. Consequently, this reduces the luminosity of the system at a given temperature. The \texttt{Simba NoAGN} result (already studied in \citealt{Robson2020}) is consistent with the results of \citet{Liang2016}.

\subsection{Total baryon, stellar and gas fractions within the galaxy groups} \label{sec: Baryon fractions}

As expected, AGN feedback has the greatest impact on the scaling relations. It lowers the gas content of the groups, thereby reducing the luminosity and increasing the entropy, with only a slight impact on the temperature. However, the addition of stellar feedback (NoFeedback to NoAGN) had a similar impact on the scaling relations, albeit less intensive. This process should also be reflected in the baryonic mass fractions in groups.  Also, as shown in Fig. \ref{fig: ST_spec_corr}, \texttt{Simba-C}, with its lower amounts of metals, requiring a `weaker' AGN feedback model, resulted in a slight but valuable improvement to the entropy, by lowering the observed gas entropy within the inner-core region of the group. In this section, we examine whether the simulated group's mass components are realistic by exploring various baryonic mass components as functions of the halo mass and compare among our different simulations and observations. This process has been extensively studied in previous literature with a specific focus on the impact of AGN on the original \texttt{Simba} simulation by comparing it with the \texttt{Simba NoAGN} simulation (see \citealt{Robson2020}), where they found that the addition of AGN feedback not only greatly impacted the mass fractions, but also resulted in the \texttt{Simba} simulation matching the observations quite well. Therefore, \texttt{Simba-C} is not expected to result in greatly differing mass fractions, but is expected to show, at most, a mild departure from the \texttt{Simba} simulation due to the alteration of the AGN feedback strength. Our focus will therefore be largely on the validation of the \texttt{Simba-C}'s mass fractions, and show the three other simulations only for comparison.

\begin{figure*}
\includegraphics[width=1.0\textwidth, trim=0.2cm 0.3cm 0.2cm 0.2cm,clip]{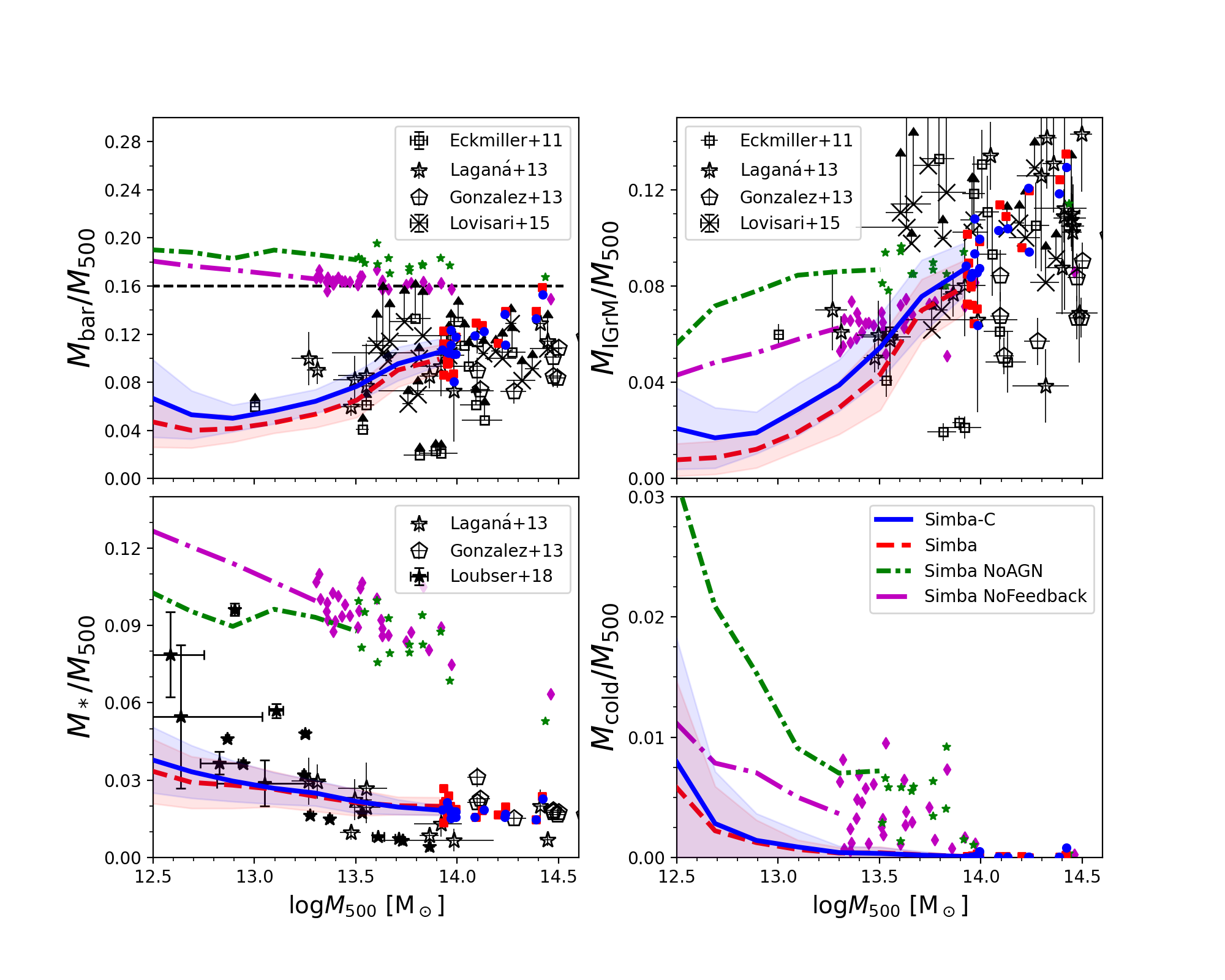}
    \caption{Stellar and gas mass fractions within $R_{500}$ for the various different simulations with their lines as described in Fig. \ref{fig: LT_spec_corr}. For comparison, observations of the following low-redshift group data are included: \citet{Eckmiller2011} (squares), \citet{Lagana2013} (open stars), \citet{Gonzales2013} (pentagons), \citet{Lovisari2015} (crosses), and \citet{Loubser2018} (filled stars). The upper left panel shows the total baryonic fraction. The black line indicates the cosmological value of the simulation, $\Omega_{b} /\Omega{m} = 0.16$. The upper right panel shows the hot gas fraction. The bottom left panel shows the stellar mass fraction. The bottom right panel shows the cold gas fraction (i.e. diffuse gas with $T < 5 \times 10^{5}\, \mathrm{ K}$ and the galactic ISM). The simulation results include stars in the galaxies and those comprising the diffuse intragroup stars (IGS) component. Only \citet{Gonzales2013} accounts for the IGS.}
    \label{fig: Mass_fraction}
\end{figure*}

Fig. \ref{fig: Mass_fraction} shows the mass fractions for four different properties: (i) The baryonic mass fraction (top left panel), (ii) the hot $T>5\times 10^{5}\,\mathrm{ K}$ diffuse IGrM gas mass fraction (top right panel), (iii) the stellar mass fraction (bottom left panel), and finally (iv) the cold IGrM gas fraction (bottom right panel) -- within $R_{500}$ for the simulated groups in our four simulations at $z=0$ versus their halo mass. For comparison, we include low redshift X-ray observational results from \citet{Eckmiller2011} using the HIFLUGC Survey, \citet{Lagana2013} using \textit{XMM-Newton, Chandra} and the Sloan Digital Sky Survey (\textit{SDSS}) observations, \citet{Gonzales2013, Lovisari2015} both using \textit{XMM-Newton} data, and \citet{Loubser2018} using Brightest Group Galaxies (BGGs) from the CLoGS sample \citep{OSullivan2012, Kolokythas2022}. It must be noted that the BGG results only provide a lower limit on the CLoGS mass scales, although it is expected that the BGG's stellar mass would dominate the CLoGS group's stellar mass. Furthermore, the CLoGS sample is chosen to have the early-type galaxies as the BGGs, which is not necessarily the case for the simulations. Therefore, the CLoGS BGGs are not a direct comparison but provide insight into the lower-mass groups. The cosmological baryon fraction assumed in these simulations of $\Omega_{b}/\Omega_{m}=0.16$ is indicated by the black dashed line.

From Fig. \ref{fig: Mass_fraction}, we see distinct groupings of models. The two simulations that include the AGN feedback produce low baryonic mass fractions in all components (also seen in figure 6 by \citealt{Robson2020}), while those lacking AGN feedback produce high baryonic contents. This split persists through every mass fraction for all mass ranges. \texttt{Simba-C} shows only a small increase in the different mass fractions compared to \texttt{Simba}, validating our hypothesis that \texttt{Simba-C} should be at most a slight departure from \texttt{Simba}. Examining the small differences between \texttt{Simba-C} and \texttt{Simba} in more detail, we note that \texttt{Simba} simulation predicts slightly lower baryonic and hot diffuse gas mass fractions than \texttt{Simba-C} at lower masses, but converges for halo masses $\log M_{500}>13.5\, \mathrm{ M_{\odot}}$. This is responsible for driving the higher entropies seen in \texttt{Simba} \citep{Oppenheimer2021}.

In summary, the primary physics module necessary to obtain realistic mass fractions is AGN feedback for \texttt{Simba}, as determined by \citet{Robson2020}. This was also found by \citet{Helden2018} using the \texttt{FABLE} simulation when AGN feedback is included, indicating that baryon fractions are a strong constraint in this process \citep[see e.g.][]{Oppenheimer2021}. Furthermore, as shown in \citet{Cui2022}, a consistent jet velocity implemented in \texttt{Simba} for the AGN feedback is more efficient at reducing the gas fractions in galaxy groups than clusters. Therefore, agreement of \texttt{Simba-C} with observations in the group regime is non-trivial and has been difficult to obtain in other models \citep[e.g.][]{Barnes2017, McCarthy2018}. \texttt{Simba-C} has an increase, albeit slight, in the mass fractions compared to \texttt{Simba}, owing to the lowering of the AGN feedback strength. This motivates us to examine the hot-gas metallicities in more detail.

\subsection{Metal enrichment of the IGrM}\label{sec: Metal enrichment of IGrM}

Given the vital role played by the chemical enrichment model in star formation through metal cooling and establishing the abundance ratios of various elements in these groups, it is interesting to delve into the chemical enrichment of these groups. 

The metal content in the intergalactic medium originates from the transport out of the ISM primarily through large-scale galactic outflows~\citep[e.g.][]{Aguirre2001, Dave2008, Oppenheimer2012, Veilleux2013}. These outflows simultaneously establish the mass-metallicity relation in galaxies \citep{Finlator2008, Dave2011, Hirschmann2013, Somerville2015, Liang2016}. However, in the group and cluster environment, it remains less clear whether these enriching outflows are driven by stellar or AGN feedback and to what extent gas stripping processes contribute.  In this context, we focus on the observed abundances and abundance ratios of silicon, iron, and oxygen. This allows us to constrain the underlying driver of IGrM enrichment and evaluate how well our models match with observations.

\begin{figure*}
\includegraphics[width=1.0\textwidth, trim=0.2cm 0.0cm 0.0cm 0.2cm,clip]{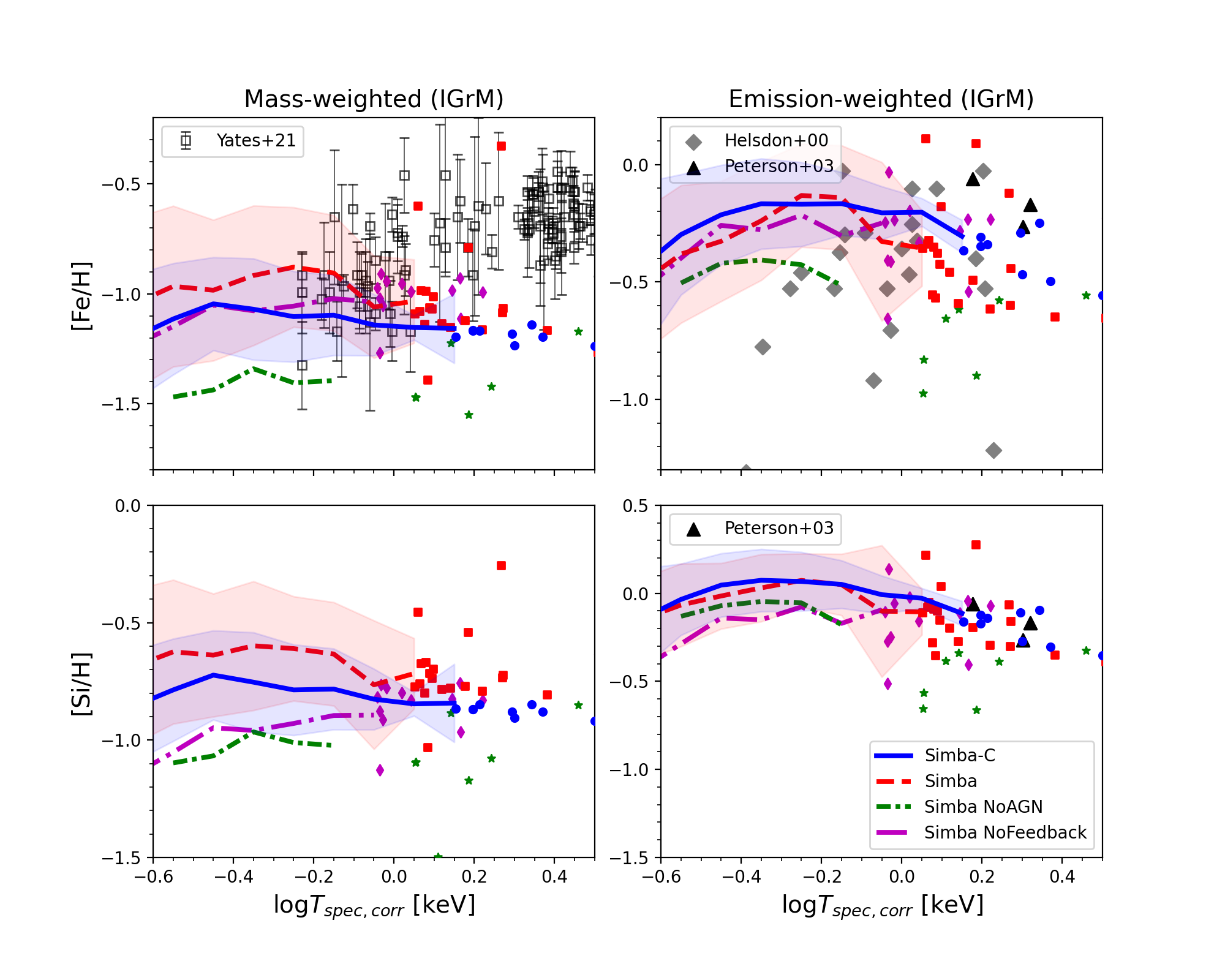}
    \caption{Global Fe (top) and Si (bottom) abundances within $R_{500}$ for the various different simulations with their lines as described in Fig. \ref{fig: LT_spec_corr}. The left column shows the mass-weighted abundances, while the right column shows the emission-weighted abundance in the IGrM. For comparison, observations of the following low redshift group data are included: \citet{Helsdon2000} (grey diamonds), \citet{Peterson2003} (black triangles) and \citet{Yates2021} (squares).} 
    \label{fig: abundance_ratios}
\end{figure*}

In Fig. \ref{fig: abundance_ratios}, we show the global Fe abundance (top row) and the global Si abundance (bottom row) for the simulated groups within $R_{500}$ in our four simulations at $z=0$ versus the halo mass. We make a distinction between the mass-weighted (left column) and emission-weighted (right column) abundances. In addition, these abundance calculations only involve hot diffuse gas. For comparison, mass-weighted X-ray [Fe/H] abundances from \citet{Yates2021} using the Mapping Nearby Galaxies at APO (MaNGA) and the Multi-Unit Spectroscopic Explorer (MUSE) observations are included. Emission-weighted X-ray [Si/H] and [Fe/H] abundances from \citet{Helsdon2000} using \textit{ROSAT} Position Sensitive Proportional Counters (PSPC) observations and \citet{Peterson2003} using \textit{XMM-Newton} data are shown in the right panels.

From Fig. \ref{fig: abundance_ratios}, it is evident that the emission-weighted abundance ratios are, on average, higher than the mass-weighted X-ray abundances. This is expected because much of the gas mass resides on the outskirts of groups where the metallicities are lower, while a significant portion of the emission originates from the central region where the metallicities are higher. This underscores the importance of creating realistic mock X-ray images, for example, using methods such as Mock Observations of X-ray Halos and Analysis (\texttt{MOXHA})~\citep{Jennings2023}, to ensure accurate comparisons with metallicity measurements. While we defer a detailed investigation of this aspect to future work, for now, we consider the emission-weighted measures as a reasonable proxy for what would be observed.

Mass-weighted abundances establish a more direct link to the underlying physical processes of group-wide enrichment. In this context, it is evident that \texttt{Simba-C} exhibits lower abundances than \texttt{Simba} in both iron and silicon, by approximately $\sim 0.2$ dex. Recall that the overall metal production rate is lower in \texttt{Simba-C}, but was re-tuned to match the galaxy mass-metallicity relation~\citep{Hough2023}. With a higher fraction of metals retained within galaxies, this worsens the differences in IGrM enrichment.

The \texttt{Simba NoAGN} model exhibits substantially lower metallicity than \texttt{Simba}. This is despite the fact that the stellar mass and hence metal production in NoAGN are substantially higher. Therefore, AGN feedback plays an major role in the ejection of metals into the IGrM. This is due to \texttt{Simba}'s AGN kinetic feedback being hydrodynamically decoupled for some time upon ejection; hence, they explicitly cannot retain ISM metals. Therefore, the impact of AGN feedback arises from quenching galaxies via the heating of ISM gas, allowing this gas, along with its associated metals, to join the hot IGrM.

In the case of NoFeedback, the IGrM metal enrichment can result only from tidal stripping. This establishes a baseline for other models, although it is essential to consider that NoFeedback produces significantly more metals overall because of the formation of a larger number of stars. 

We now focus on the emission-weighted X-ray Fe and Si abundances in the right panels. Generally, most models show an abundance of [Fe/H] of about $\sim 0.2-0.4$, with the NoAGN model slightly below this. However, considering that both the models and the observations show a large scatter in the [Fe/H] abundance, we cannot draw conclusive conclusions. [Si/H] predictions in all models appear to be broadly similar to the observations, but the data do not extend into the group regime to allow direct comparisons. Interestingly, there is not much difference between the original \texttt{Simba} simulation and the \texttt{Simba-C} simulation, or even the simulations without feedback. This is in contrast with the mass-weighted abundances, illustrating the high bias when measuring metallicities only from the X-ray-emitting gas and cautioning against over-interpretation of such data in terms of metal formation mechanisms and timescales. The results can be reconciled by noting that much of the X-ray emission comes from the central region, so potentially the central metallicities could be similar even if the overall mass-weighted ones (dominated by mass in the outskirts) are different. In future work, we will examine group metal profiles.

\begin{figure*}
\includegraphics[width=1.0\textwidth, trim=1.4cm 1.2cm 1.0cm 1.6cm,clip]{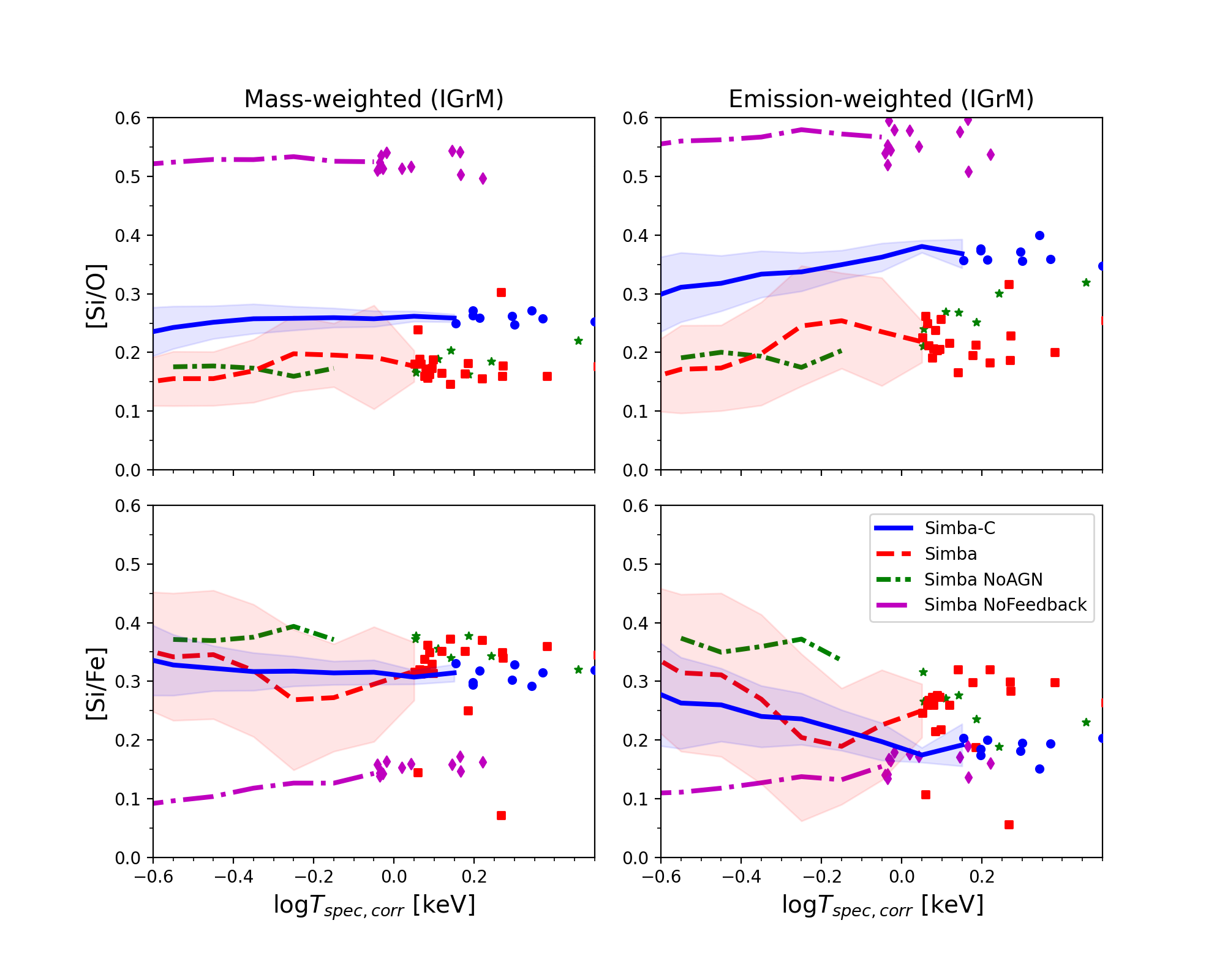}
    \caption{Global [Si/O] (top) and [Si/Fe] (bottom) abundance ratios within $R_{500}$ for the various different simulations with their lines as described in Fig. \ref{fig: LT_spec_corr}. The left column shows the mass-weighted abundances, while the right column shows the emission-weighted abundance in the IGrM.}
    \label{fig: abundance_ratios_O}
\end{figure*}

Fig. \ref{fig: abundance_ratios_O} is similar to Fig. \ref{fig: abundance_ratios}, but shows the global abundance ratios [Si/O] (top row) and [Si/Fe] (bottom row) instead of the chemical abundances.\footnote{Other notable group and cluster abundance ratios that will be looked at in future work include [Ni/Fe], [Ne/Fe], [O/Fe], [Mg/O], [Ne/O], and O (solar). Observational results for these ratios have recently been shown in \citet{Fukushima2023}.} These abundance ratios are scaled to the solar level to \citet{Anders1989}.

From Fig. \ref{fig: abundance_ratios_O}, we observe notable differences between the four simulations. First, unlike in Fig. \ref{fig: abundance_ratios}, we do not see significant differences between the mass-weighted and emission-weighted calculations. Second, three clear trends emerge: (i) The \texttt{Simba-C} simulation have a significantly higher [Si/O] abundance ratios with respect to \texttt{Simba}. The \texttt{Simba} simulation is the only simulation that matches \texttt{Simba-C}'s [Si/Fe] trend; however, \texttt{Simba} has a large $1\sigma$-error range that covers the entire \texttt{Simba-C}'s tight spread. Therefore, the matching of \texttt{Simba} in this case could only be the result of the large error region. (ii) AGN feedback seems to play a minor role in these abundances with its inclusion in \texttt{Simba} having no effect on [Si/O] and only a slight impact on [Si/Fe]. (iii) All simulations have matching slopes/trends, albeit at different values. 

From these differences, it is clear that with the introduction of a stellar feedback system and its chemical enrichment process (NoFeedback $\rightarrow$ NoAGN), a specific trend/ lope is obtained as a function of temperature. The slope height was only significantly altered again with the introduction of the updated chemical enrichment model and the stellar feedback system. Combining this with the fact that the abundance ratios seem to be insensitive to AGN feedback, it shows that it is crucial that the simulation contains an accurate stellar feedback and chemical enrichment model for the creation of realistic groups and clusters.

\section{Redshift evolution of various physical properties as a function of group temperature}\label{sec: Redshift evolution}

In this section, we briefly study the evolution of the $1\,\mathrm{keV}$-temperature groups at all redshifts. We track a specific type of halo as a function of redshift ($z=2$ to $0$) for all four of our simulations. We discuss the same plots as in the previous section (Sec. \ref{sec: scaling relations}). However, since this is an evolution study of very particular simulated galaxies groups, we will not show any observations, which are only for $z\approx 0$. Furthermore, this section only concentrates on available trends to see if interesting topics can emerge for future studies, and our main goal of determining how the \texttt{Simba-C} simulation with its updated physics modules compares with observations and to \texttt{Simba} remains.

\subsection{Scaling relations}\label{sec: redshift scaling}

\begin{figure}
\includegraphics[width=\columnwidth, trim=0.15cm 0.2cm 0.1cm 0.15cm,clip]{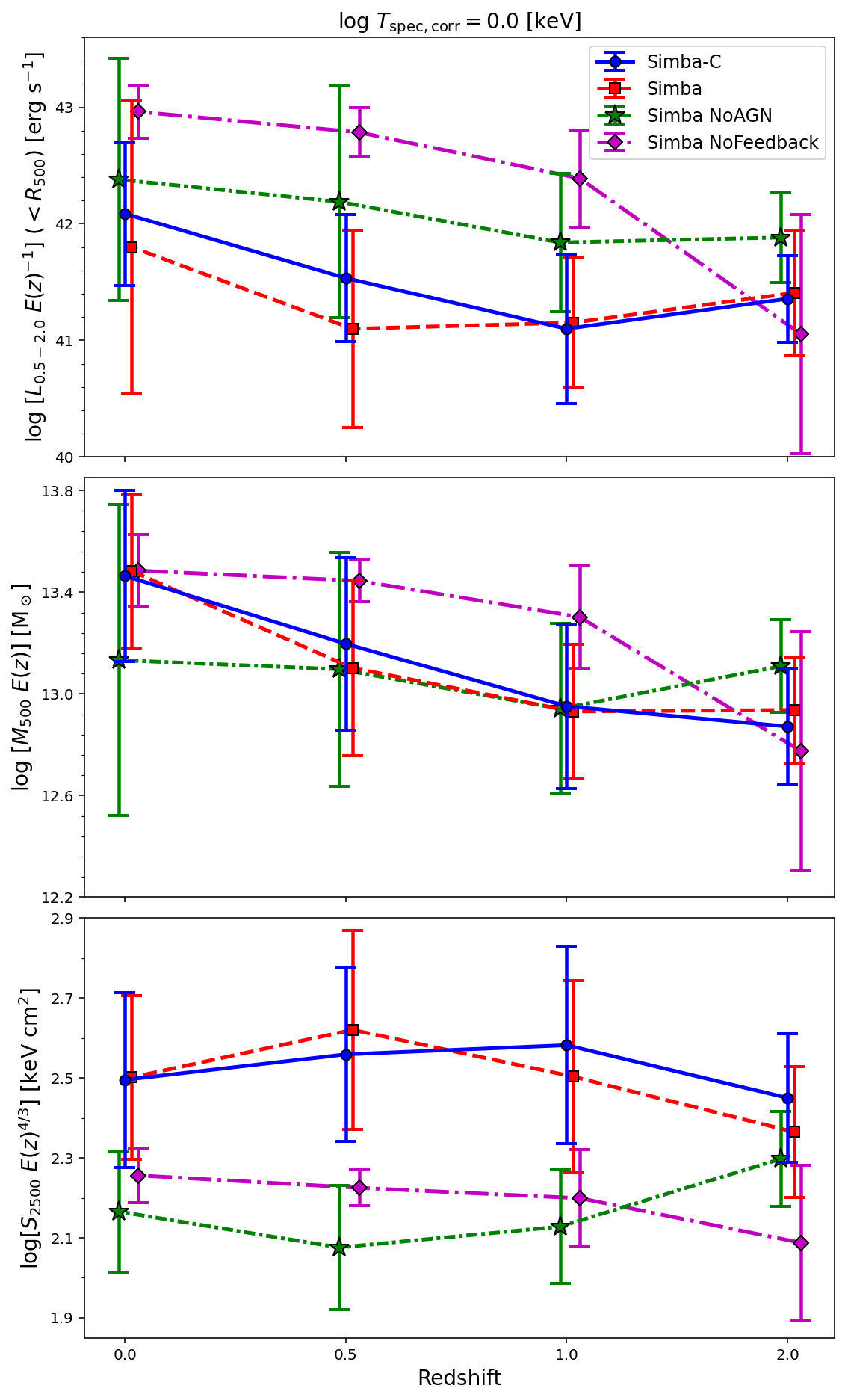}
    \caption{The average: a) $L_{X,0.5-2.0}$-values (upper panel) and b) $M_{500}$-value (middle panel) both within $R_{500}$, as well as the c) gas entropy $S_{2500}$ within $R_{2500}$ (bottom panel), for the simulated group haloes with $T_{spec,corr}=1\,\mathrm{keV}$ over redshift. The $1\sigma$-error bars are shown. The simulations included for comparison are: \texttt{Simba-C} (blue circle/solid line), \texttt{Simba} (red square dashed line), \texttt{Simba NoAGN} (green star/tight dot-dashed line), and \texttt{Simba NoFeedback} (purple diamond/loosely dot-dashed line). We include a small offset at each redshift to separate the error bars for visibility.}
    \label{fig: LT_spec_corr_z}
\end{figure}

Here, we present the three scaling relations in specifically chosen temperature bins with equally spaced bin sizes of $\log T_{spec} = 0.075\,\mathrm{keV}$ ($\log M_{500} = 0.208\,\mathrm{M_{\odot}}$), at redshifts $z=2,\,1,\,0.5$ and $z=0$. Fig. \ref{fig: LT_spec_corr_z} shows at $T_{spec,corr}=1\,\mathrm{ keV}$ values (i) upper panel - the group luminosity $L_{X,0.5-2.0}$, (ii) middle panel - the group mass $M_{500}$, and (iii) bottom panel - the inner-group region gas entropy $S_{2500}$.\footnote{The $S_{500}$ evolutionary track has been omitted here due to the small difference between \texttt{Simba-C} and \texttt{Simba} seen in Fig. \ref{fig: ST_spec_corr}.} Table \ref{tab: halo amounts temp bins} shows the number of haloes in the $1\,\mathrm{keV}$ group temperature bin used in Fig. \ref{fig: LT_spec_corr_z}. Due to the smaller box volume of the \texttt{Simba NoAGN} and \texttt{Simba NoFeedback} the 10 halo limit had to be lowered to 8 haloes to produce complete evolutionary tracks only for these two simulations. This reduces the statistical reliability for these two simulations, but it is a necessary modification.

\begin{table}
\centering
\begin{tabular}{|c|c|c|c|c|}
\hline
Simulation & $z=0$ & $z=0.5$ & $z=1$ & $z=2$\\
\hline
\texttt{Simba-C} ($100\, \mathrm{Mpc} \, h^{-1}$)& 46 & 84 & 174 & 213 \\
\texttt{Simba} ($100\, \mathrm{Mpc} \, h^{-1}$)& 46 & 103 & 213 & 154 \\
\texttt{Simba NoAGN} ($50\, \mathrm{Mpc} \, h^{-1}$)& 9 & 17 & 26 & 26 \\
\texttt{Simba NoFeedback} ($50\, \mathrm{Mpc} \, h^{-1}$)& 14 & 8 & 9 & 21 \\
\hline
\end{tabular}
\caption{Amount of haloes in the $1\,\mathrm{keV}$ group temperature bin used in Fig. \ref{fig: LT_spec_corr_z}.}
\label{tab: halo amounts temp bins}
\end{table}

We present these results for each of our four simulations: (i) The full \texttt{Simba-C} simulation (blue circle solid line), (ii) the original \texttt{Simba} simulation (red square dashed line), (iii) the \texttt{Simba NoAGN} simulation (green star tight dot-dashed line), and finally (iv) the \texttt{Simba NoFeedback} simulation (purple diamond loosely dot-dashed line). In addition, we show the $1\sigma$-deviation for each calculated quantity at every redshift for all four of our simulations, indicated by the error bars. We applied a small offset at each redshift to separate the error bars, making it easier to distinguish them in the plot. We use this approach throughout the entirety of Sec. \ref{sec: Redshift evolution}. We consider the following redshifts for each plot: $z=2$ (Cosmic Noon -- highest star formation rate), $z=1$, $z=0.5$ (start of the accelerated cosmic expansion), and $z=0$ (present-day).

From Fig. \ref{fig: LT_spec_corr_z} (upper panel), we observe that without feedback, the evolution of $L_{X,0.5-2.0}$ for the $1\,\mathrm{ keV}$ groups consistently exhibits increasing luminosity, resulting, on average, in the brightest haloes. However, when feedback (first stellar and then AGN) is included, the luminosity is reduced by almost an order of magnitude by $z=1$, and then followed by a continued lowered luminosity up to $z=0$ for \texttt{Simba-C} and \texttt{Simba}. Interestingly, the increasing luminosity appears to start only from $z=1$ to $0$, while without stellar feedback, it increases throughout the evolution between $z=2$ and $z=0$. However, when the large errors are taken into account, only the general increasing luminosity as a function of redshift trend and the fact that feedback lowers the luminosity in the galaxy groups' early evolution are significant. The differences between each individual simulation provide only potential nonsignificant patterns, e.g. stellar feedback impacting the period when the group's luminosity begins to evolve.

From Fig. \ref{fig: LT_spec_corr_z} (middle panel), we observe that the evolution of the $M_{500}$ for the $1\,\mathrm{ keV}$ groups in our four simulations is similar to the trends found in the $L_{X,0.5-2.0}-T_{spec,corr}$ plot (upper panel). The masses of the haloes increase with time, similar to the increase of the luminosity with time. However, the four simulations yielded more closely aligned trends, particularly between \texttt{Simba-C} and the original \texttt{Simba} simulation, which resulted in nearly identical evolutionary tracks for $M_{500}-T_{spec,corr}$. Therefore, the \texttt{Chem5} model plays no role in the evolution of this scaling relation.

From Fig. \ref{fig: LT_spec_corr_z} (bottom panel), we observe that the gas entropy in the haloes' inner core region -- $R_{2500}$ -- for the $1\,\mathrm{ keV}$ groups seems to have minimal evolution. Furthermore, AGN feedback, as expected, appears to have the largest impact on the outcome of the gas entropy, resulting in the only significant difference between the four simulation's entropy. The only other interesting effect is that both \texttt{Simba} and \texttt{Simba-C} appear to experience a decrease in gas entropy between $z=0.5$ and $z=0$ for the former and between $z=1$ and $z=0$ for the latter in the inner core region. This change is very small, and when the large errors are taken into account, this effect is negligible in this study but deserves a more detailed investigation to determine the origin of this effect.

\subsection{Evolution of physical properties within galaxy groups} \label{sec: Baryon redshift fractions}

\begin{figure*}
\includegraphics[width=1.0\textwidth, trim=0.25cm 0.25cm 0.1cm 0.25cm,clip]{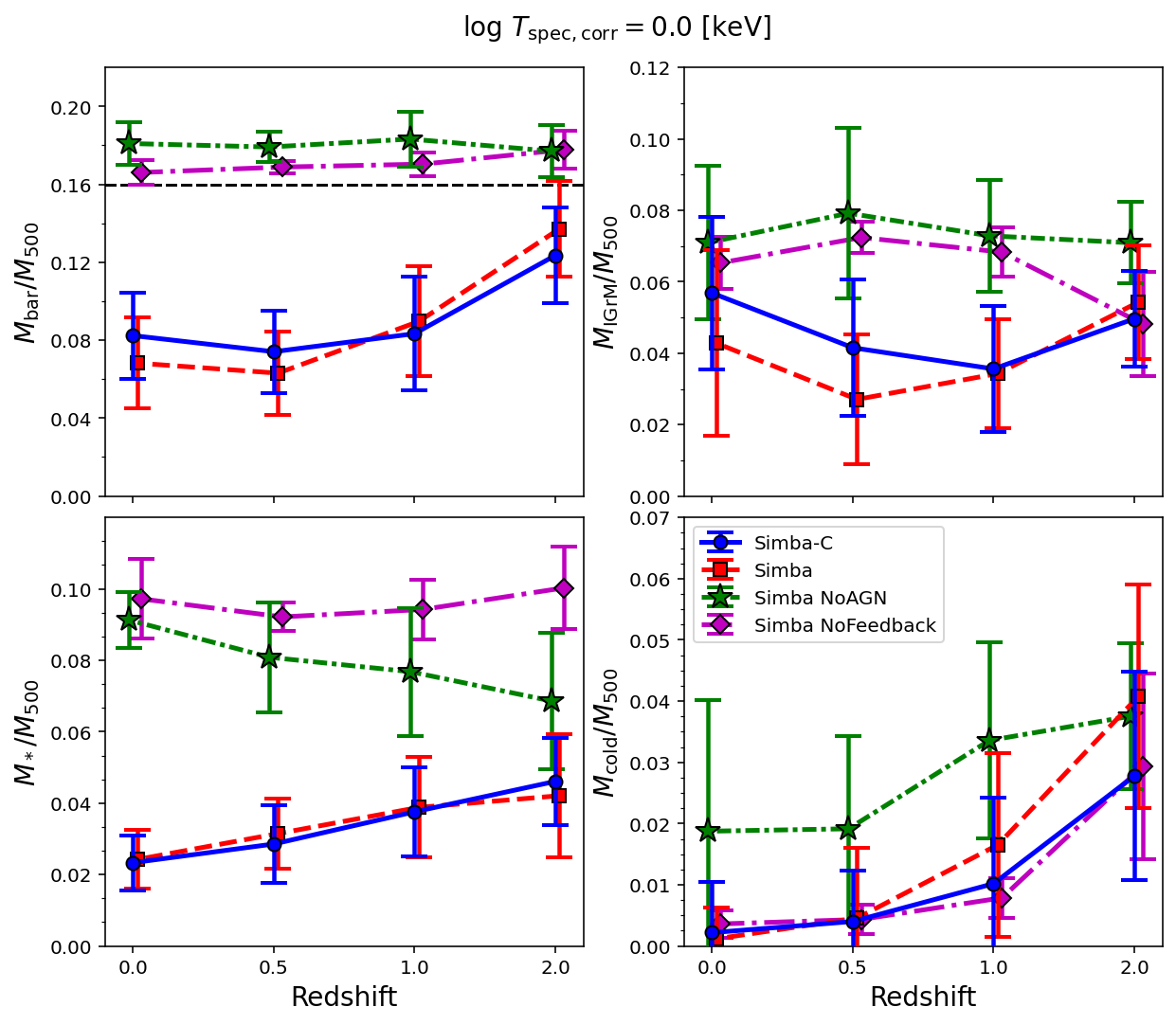}
    \caption{The average stellar and gas mass fractions within $R_{500}$ for the simulated group haloes with $\log T_{spec,corr}=0\, \mathrm{ keV}$ over redshift. The $1\sigma$-error bars are shown. The simulations included for comparisons are shown with the same lines as described in Fig. \ref{fig: LT_spec_corr_z}. The upper left panel shows the total baryonic fraction. The black line indicates the simulation's cosmological value, $\Omega_{b} /\Omega_{m} = 0.16$. The upper right panel shows the hot gas fraction. The bottom left panel shows the stellar mass fraction. The bottom right panel shows the cold gas fraction (i.e. diffuse gas with $T < 5 \times 10^{5}\, \mathrm{ K}$ and the galactic ISM).}
    \label{fig: Mass_fractions_z}
\end{figure*}

\begin{figure*}
\includegraphics[width=1.1\textwidth, trim=1.4cm 1.2cm 1.0cm 0.4cm,clip]{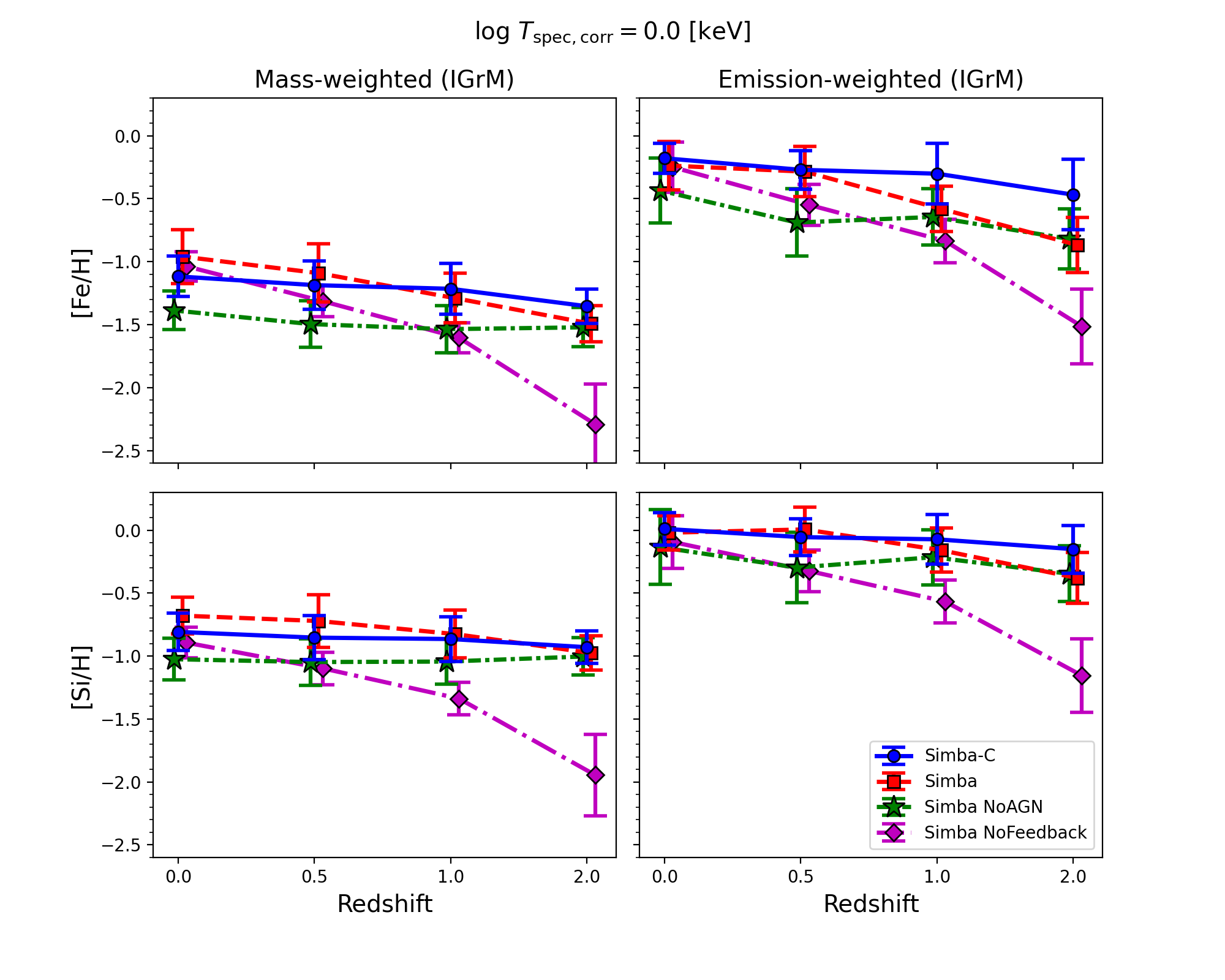}
    \caption{Global Fe (top) and Si (bottom) abundances within $R_{500}$ for the simulated group haloes with $T_{spec,corr}=1\, \mathrm{ keV}$ over redshift. The $1\sigma$-error bars are shown. The simulations included for comparisons are shown with the same lines as described in Fig. \ref{fig: LT_spec_corr_z}. The left column shows the mass-weighted abundances, while the right column shows the emission-weighted abundances in the IGrM.} 
    \label{fig: abundance_ratios_z}
\end{figure*}

\begin{figure*}
\includegraphics[width=1.1\textwidth, trim=1.4cm 1.2cm 1.0cm 0.4cm,clip]{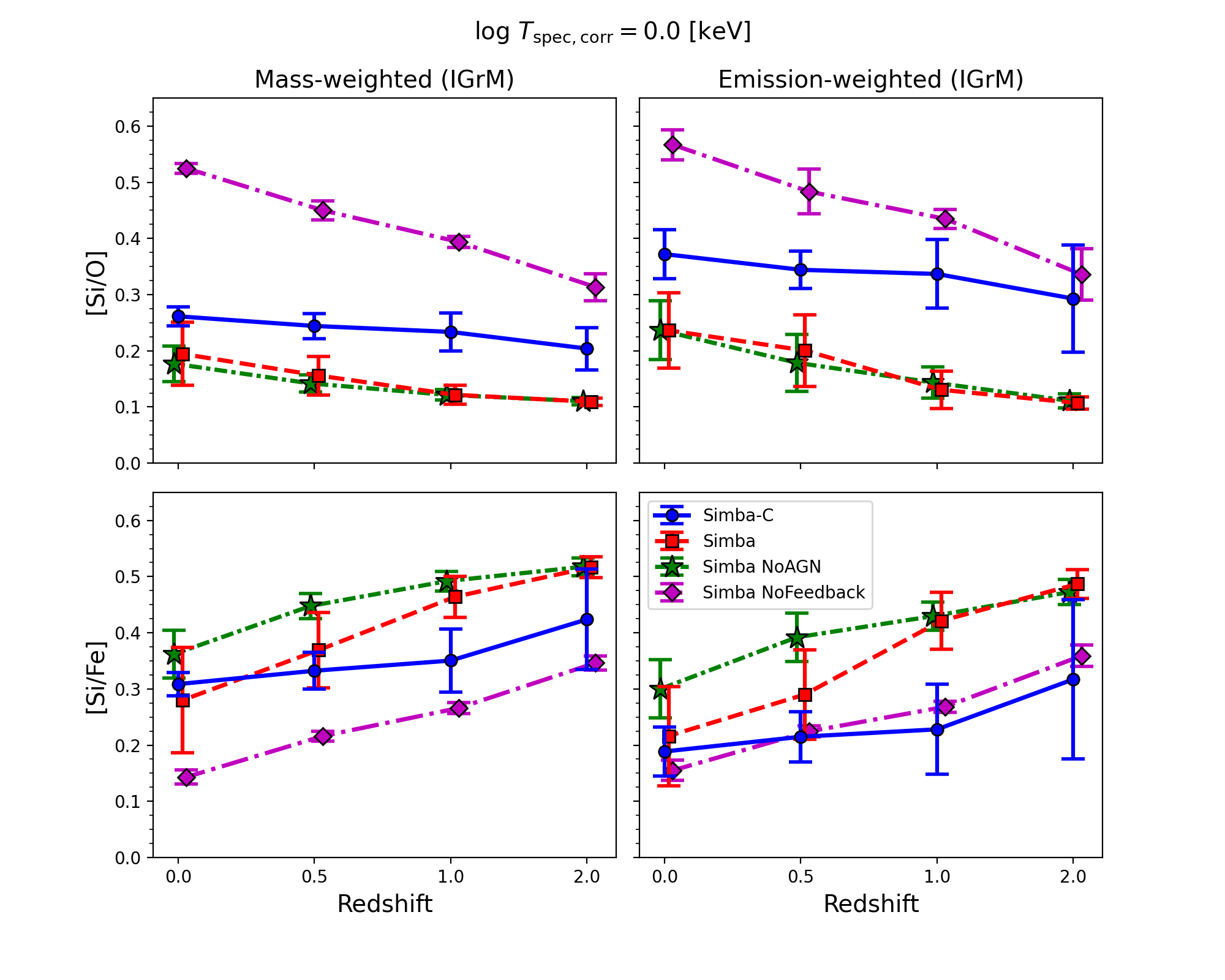}
    \caption{Global Si-O (top) and Si-Fe (bottom) abundance ratios within $R_{500}$ for the simulated group haloes with $T_{spec,corr}=1\,\mathrm{ keV}$ over redshift. The $1\sigma$-error bars are shown. The simulations included for comparisons are shown with the same lines as described in Fig. \ref{fig: LT_spec_corr_z}. The left column shows the mass-weighted abundance ratios, while the right column shows the emission-weighted abundance ratios in the IGrM.}
    \label{fig: Abundance_ratio_O_z}
\end{figure*}

Similarly to the previous section, we discuss the evolution of the various galaxy group properties; however, here we focus on the evolution of the physical properties that govern the group structure. We show the following plots in the specifically chosen $T_{spec,corr}=1\, \mathrm{keV}$ group temperature bin: (i) mass fractions are shown in Fig. \ref{fig: Mass_fractions_z}, (ii) global Fe and Si abundances are shown in Fig. \ref{fig: abundance_ratios_z} , and finally (iii) global abundance ratios [Si/O] and [Si/Fe] are shown in Fig. \ref{fig: Abundance_ratio_O_z}. Table \ref{tab: halo amounts temp bins} shows the number of haloes in the $1\,\mathrm{keV}$ group temperature bin used in Figs. \ref{fig: Mass_fractions_z} - \ref{fig: Abundance_ratio_O_z}.

From Fig. \ref{fig: Mass_fractions_z}, we note that the trends appear to be similar to the group mass fractions in Fig. \ref{fig: Mass_fraction}. However, these trends show only the evolution of a specific type of galaxy group with a halo spectroscopic temperature of $T_{spec,corr}=1\, \mathrm{ keV}$. Interestingly, between \texttt{Simba-C} and \texttt{Simba} there are minor notable differences that arise in the evolution of the mass fractions. Specifically, it seems that \texttt{Simba-C} has more hot diffuse gas at $z=0$, while \texttt{Simba} has more cold gas at $z=2$. This results in a slight difference in the baryonic mass over the evolution period, but with virtually no difference in the stellar mass.

Since AGN feedback plays an important role in mass fraction calculations, these minor differences can be a direct result of the re-calibrated AGN feedback strength in \texttt{Simba-C}. The differences between \texttt{Simba NoFeedback} and \texttt{Simba NoAGN} simulations gives further indication that stellar feedback can affect the fraction of the mass component and to some extent the evolution and should therefore be taken into account, but AGN feedback remains the primary contributor to the observed differences seen in the mass fractions, as expected.

From Fig. \ref{fig: abundance_ratios_z}, we note that the evolution of the global Si and Fe abundances within these galaxy groups is remarkably similar for three of the simulations. The only major exception to these trends appears to be the \texttt{Simba NoFeedback} simulation, which experiences a substantial increase in both these elements between $z=2$ and $z=0$; however, this could be the result of the low number of haloes in $50\, \mathrm{Mpc} \, h^{-1}$. The other three simulations also show an increase in Fe and Si, although only slightly. This increase is expected to occur as a result of the creation of new metals through stellar feedback. 

If the increase experienced by the \texttt{Simba NoFeedback} is not related to the low amount of haloes, it is interesting to note that even without stellar feedback this simulation is still able to obtain similar global Si and Fe abundances at $z=0$ solely through stellar evolution and an abundance of star formation from the cooling gas. Therefore, the inclusion of metals with star formation through cooling will still evolve these abundances, regardless of the accompanying stellar feedback model. Hence, the yields should be as accurate as possible to allow the accompanying stellar feedback model to obtain the resulting abundance trends at the correct stage in the galaxy group's evolution. This again motivates the need for the newly updated chemical enrichment model when studying the chemical evolution of the systems.

Fig. \ref{fig: Abundance_ratio_O_z} shows that the global abundance ratios evolve significantly and that they are very sensitive to the stellar feedback physics models included in the simulations. This is one of our most significant differences between the four simulations on group scales. We note three interesting phenomena within the ratios and weighted temperatures for the galaxy groups with $T_{spec,corr}=1\,\mathrm{ keV}$. First, the trends/slopes remain relatively consistent regardless of the simulation. Only the values of the abundance ratios differ (either all of them systematically increase or decrease). Second, in both abundance ratios, a $\gtrsim 0.2\, \mathrm{ dex}$ decrease/increase is obtained as a result of the addition of the instantaneous recycling of metals approximation and its stellar feedback model. This is then reversed by $\sim 0.1$ dex with the updated stellar feedback and chemical enrichment model in \texttt{Simba-C}. Third, it seems that AGN feedback plays a minimal role in the evolution of the abundance ratios, especially for [Si/O], confirming Fig. \ref{fig: abundance_ratios_O}'s conclusion.

If we focus only on the average, since the $1\sigma$-error range indicates a non-significant difference between \texttt{Simba-C} and \texttt{Simba}, we find an interesting phenomenon between these two simulations. On average, \texttt{Simba-C} obtained a higher abundance of Si compared to the abundance of O, as seen in the increase in the abundance ratio of [Si/O] (Fig. \ref{fig: abundance_ratios_O}) compared to \texttt{Simba}. On the other hand, they both obtained a similar [Si/Fe] abundance ratio at $z=0$ (also in Fig. \ref{fig: abundance_ratios_O}). This indicates that Si and Fe scales similarly between the two simulations. Therefore, we have an increase in the abundance of Si and Fe relative to the abundance of O in \texttt{Simba-C}. This corresponds to the frequently used [$\alpha$/Fe] trend \citep{Wallerstein1962, Kobayashi2020a, Kobayashi2023}. This trend has a plateau for high [$\alpha$/Fe] abundance ratios at [Fe/H]$<-1$ values and then decreases to [$\alpha$/Fe]$\sim 0$ after [Fe/H]$\sim -1$ (owing to an increase in Fe from SNe Ia). \texttt{Simba-C} can successfully reproduce this pattern unlike \texttt{Simba}, as shown in \citet{Hough2023}. However, the interplay between the Si, O, and Fe abundances is complex and may not be fully understood from these graphs. A possible explanation for the lowering of O relative to Si and Fe could be the result of the introduction of `failed' SNe, as shown in \citet{Kobayashi2020a}, where all three of these elements were affected.

From the above discussion, we can conclude that to obtain the most realistic abundance ratios, an accurate chemical enrichment and its corresponding stellar feedback model are crucial.

\section{Summary}\label{sec: Conclusions}

In this paper, we examine the halo and galaxy group X-ray properties in detail for the cosmological simulation known as \texttt{Simba-C}, the most recent and up-to-date version of \texttt{Simba}. A significant fraction of the baryons within these galaxy groups exist in the form of hot diffuse gas, enabling the study of galaxy groups through X-ray observations.

To identify the haloes containing the galaxy groups within our simulation, we employed the Amiga Halo Finder to generate a catalogue containing information about gas, dark matter, and star particles, along with their corresponding host galaxies, located within each halo. These catalogues were then utilized to analyze the halo X-ray properties, including radius, temperature (both $T_{X}$ and $T_{spec,corr}$), luminosity, mass, entropy, and metallicity, using the X-ray property of the IntraGroup Medium Python package (XIGrM).

Utilizing these catalogues, we presented and discussed general halo properties such as the halo mass function, the galaxy stellar mass function, and the virial mass and X-ray luminosity as a function of $T_{X}$.

Furthermore, due to the complexities surrounding the various feedback mechanisms (i.e. stellar and AGN feedback) and metal content (i.e., metal yields) and their influence on simulations, we used different versions of the \texttt{Simba} simulation in an attempt to understand how each physics module's implementation influenced the X-ray properties, with a specific focus on the updated chemical enrichment and stellar feedback physics. These simulations include two $100\, \mathrm{Mpc} \, h^{-1}$ box simulations: (i) The original published \texttt{Simba} simulation, with an instantaneous recycling of the metals model approximation, an AGN feedback model, and a dust model. (ii) Our main result, the complete \texttt{Simba-C} simulation with its updated chemical enrichment, a self-consistent stellar feedback model, a recalibrated AGN feedback strength, and the reintegrated dust model from \texttt{Simba}. These two simulations are accompanied by two $50\, \mathrm{Mpc} \, h^{-1}$ box simulations for comparison: (iii) A no feedback \texttt{Simba} simulation, and a (iv) \texttt{Simba} simulation with only the simplistic instantaneous recycling of the metals model.

In Sec. \ref{sec: scaling relations}, we presented the present-day ($z=0$) X-ray scaling relations, namely $L_{X,0.5-2.0}-T_{spec,corr}$, $M_{500}-T_{spec,corr}$, and $S_{500/2500} - T_{spec,corr}$ (Figs. \ref{fig: LT_spec_corr}, \ref{fig: MT_spec_corr}, and \ref{fig: ST_spec_corr}, respectively). We compared the scaling relations with various low-redshift X-ray observations. The first notable result is that the complete \texttt{Simba-C} simulation (our main simulation) appears to be the most consistent in matching the observations of the three scaling relations. This demonstrates that the new chemical enrichment, with its accompanying stellar feedback model, as well as the re-calibration process regarding the AGN feedback strength, is a necessary addition to the \texttt{Simba} simulation.

Second, as expected, the AGN feedback is the most important mechanism for obtaining realistic scaling relations in these simulations, matching the findings from \citet{Robson2020}, while the dust had minimal effects on the outcome of these relations. \texttt{Simba-C} reduces some of the overcorrected halo X-ray properties that came about with the introduction of AGN feedback (see Fig. \ref{fig: ST_spec_corr}). This could be due to the \texttt{Chem5} model, which originally produced fewer metals, leading to a weak metal cooling function, which resulted in the recalibration of the AGN feedback strength. 

Physical properties, namely, mass fractions and abundance ratios, also showed that AGN feedback played an important role in determining physical properties (Figs. \ref{fig: Mass_fraction}, \ref{fig: abundance_ratios}, and \ref{fig: abundance_ratios_O}, respectively). In fact, for the mass fractions, AGN feedback is the only necessary physics module to obtain simulations that can match the observations. From the abundance ratios, we observed that the \texttt{Simba-C} simulation resulted in a [Si/O] trend that differs from \texttt{Simba}. Interestingly, only the introduction of stellar feedback (NoFeedback to NoAGN) or the update to the chemical enrichment model (\texttt{Simba} to \texttt{Simba-C}) changed the abundance ratios, with the latest change partially reverting some of the changes obtained in the first update.

In Sec. \ref{sec: redshift scaling}, we presented the same properties; however, we showed the evolution of the $1\,\mathrm{keV}$ temperature groups at redshifts $z=2$, $z=1$, $z=0.5$, and $z=0$ (Figs. \ref{fig: LT_spec_corr_z} to \ref{fig: Abundance_ratio_O_z}). In most cases, we reach the same conclusions; for instance, AGN feedback being the largest contributor to the differences shown in the simulations. However, the global [Si/O] and [Si/Fe] results (Fig. \ref{fig: Abundance_ratio_O_z}) showed that the abundance ratios are sensitive to stellar feedback. Furthermore, the metal yields of the \texttt{Chem5} model produced an increase in the abundance of Si and Fe, relative to O (an $\alpha$-element). This is expected due to the pattern emerging from the frequently-used [$\alpha$/Fe] ratios, which \texttt{Simba-C} can successfully reproduce, unlike \texttt{Simba}. Therefore, even though AGN feedback is crucial for the simulation to obtain realistic galaxy/galaxy groups, an accurate stellar feedback and its chemical enrichment model are needed to produce realistic abundance ratios. 

Lastly, two minor interesting patterns were noted: (i) \texttt{Simba-C} did not change the already correctly modelled $M_{500}-T_{spec,corr}$ scaling relation and the mass fractions. Both of which \texttt{Simba} already managed to match the observations. (ii) \texttt{Simba-C} also has more hot diffuse gas at $z = 0$, while \texttt{Simba} has more cold gas at $z = 2$, slightly impacting the baryonic mass during the evolution period, but with virtually no difference in the stellar mass.

Future work will include a follow-up study based on the X-ray profiles for each galaxy group property with comparisons to the recent findings of \citet{Altamura2023}, where they found that \texttt{EAGLE}-like simulation models do not solve the entropy core problem by studying the $S/S_{500}$ profiles, which were revealed to be to flat (\textit{Padawer-Blatt et al.} in prep.). This would contextualize our improved gas entropy $S_{500}$ results, since the original \texttt{Simba} simulation also showed a flat gas entropy profile. We are also investigating a follow-up study based on the impact that the \texttt{Chem5} module has on simulated clusters and their scaling relationships and global properties. This will allow us to determine the \texttt{Chem5} model's impact on various scales, starting with individual stellar populations to galaxies, groups of galaxies, and clusters.

\section*{Acknowledgements}

This work is based on research supported in part by the National Research Foundation of South Africa (NRF Grant Number: 146053). The simulations and analysis reported in this article were enabled by HPC resources provided by the Digital Research Alliance of Canada (alliancecan.ca) award to AB. RH also acknowledges the \texttt{Simba} collaboration for the use of the simulation. RH also acknowledges the Royal Society of Science travel grant that allowed for in-person collaboration at the University of Edinburgh, Scotland, to further this study. AB and DR acknowledge the support of the Natural Sciences and Engineering Research Council of Canada (NSERC) through its Discovery Grant program. Additionally, DR acknowledges support from NSERC through a Canada Graduate Scholarship (funding reference number 534263), and AB acknowledges support from the Infosys Foundation via an endowed Infosys Visiting Chair Professorship at the Indian Institute of Science. Additionally, AB acknowledges the l{\fontencoding{T4}\selectfont\M{e}}\'{k}$^{\rm w}${\fontencoding{T4}\selectfont\M{e}\m{n}\M{e}}n peoples on whose traditional territory the University of Victoria stands, and the Songhees, Equimalt and WS\'{A}NE\'{C} peoples whose historical relationships with the land continue to this day. WC is supported by the STFC AGP Grant ST/V000594/1, the Atracci\'{o}n de Talento Contract no. 2020-T1/TIC-19882 was granted by the Comunidad de Madrid in Spain, and the science research grants were from the China Manned Space Project. He also thanks the Ministerio de Ciencia e Innovación (Spain) for financial support under Project grant PID2021-122603NB-C21 and HORIZON EUROPE Marie Sklodowska-Curie Actions for supporting the LACEGAL-III project with grant number 101086388. CK acknowledges funding from the UK Science and Technology Facility Council through grant ST/R000905/1 and ST/V000632/1. The Flatiron Institute is supported by the Simons Foundation. Any opinion, finding, and conclusion or recommendation expressed in this material is that of the author(s), and the NRF does not accept any liability in this regard.

%%%%%%%%%%%%%%%%%%%%%%%%%%%%%%%%%%%%%%%%%%%%%%%%%%
\section*{Data Availability}
The \texttt{Simba-C} simulation data underlying this article will be shared on reasonable request to the corresponding author. The published \texttt{Simba} simulation \citep{Dave2019} is available on the \texttt{Simba} university repository at \url{http://Simba.roe.ac.uk/}.



\bibliographystyle{mnras}
\bibliography{example} % if your BibTeX file is called example.bib

%\begin{thebibliography}{99}
%\end{thebibliography}

% Don't change these lines
\bsp	% typesetting comment
\label{lastpage}
\end{document}